\documentclass[prd,amsmath,showpacs]{revtex4}
\usepackage{bm}
\usepackage{amsmath, amsthm, amssymb}
\usepackage{bm}

\newcommand{\beq}{\begin{equation}}
\newcommand{\eeq}{\end{equation}}
\newcommand{\beqa}{\begin{eqnarray}}
\newcommand{\eeqa}{\end{eqnarray}}
\newcommand{\beqar}{\begin{eqnarray*}}
\newcommand{\eeqar}{\end{eqnarray*}}

\newtheorem{thm}{Theorem}[section]
\newtheorem{cor}[thm]{Corollary}

\newcommand\munu{\ensuremath{{\mu\nu}}}

\usepackage{bm}

\begin{document}

\title{Electrically charged fluids with pressure in Newtonian
gravitation and general relativity in $\mathbf{ d}$ spacetime
dimensions: theorems and results for Weyl type systems}

\author{Jos\'e P. S. Lemos}
\affiliation{Centro Multidisciplinar de Astrof\'{\i}sica -- CENTRA,
Departamento de F\'{\i}sica, 
Instituto Superior T\'ecnico - IST,
Universidade T\'ecnica de Lisboa - UTL,
Av. Rovisco Pais 1, 1049-001 Lisboa, Portugal,
email: lemos@fisica.ist.utl.pt, \\ \&
\\
Coordenadoria de Astronomia e Astrof\'{\i}sica, Observat\'orio
Nacional-MCT, Rua General Jos\'e Cristino 77, 20921-400, Rio de Janeiro,
Brazil, email: lemos@on.br}
\author{Vilson T. Zanchin}
\affiliation{Centro de Ci\^encias Naturais e Humanas, Universidade Federal
do ABC, \\ Rua Santa Ad\'elia, 166, 09210-170, Santo Andr\'e, Brazil,
email: zanchin@ufabc.edu.br}

\begin{abstract}

Previous theorems concerning Weyl type systems, including
Majumdar-Papapetrou systems, are generalized in two ways, namely, we
take these theorems into d spacetime dimensions (${\rm d}\geq4$), and
we also consider the very interesting Weyl-Guilfoyle systems, i.e.,
general relativistic charged fluids with nonzero pressure.  In
particular within Newton-Coulomb theory of charged gravitating fluids,
a theorem by Bonnor (1980) in three-dimensional space is generalized
to arbitrary $({\rm d}-1)>3$ space dimensions. Then, we prove a new
theorem for charged gravitating fluid systems in which we find the
condition that the charge density and the matter density should obey.
Within general relativity coupled to charged dust fluids, a theorem by De
and Raychaudhuri (1968) in four-dimensional spacetimes in rendered
into arbitrary ${\rm d}>4$ dimensions. Then a theorem, new in ${\rm
d}=4$ and ${\rm d}>4$ dimensions, for Weyl-Guilfoyle systems, is
stated and proved, in which we find the condition that the charge
density, the matter density, the pressure, and the electromagnetic
energy density should obey.  This theorem comprises, as particular
cases, a theorem by Gautreau and Hoffman (1973) and results in four
dimensions by Guilfoyle (1999). Upon connection of an interior charged
solution to an exterior Tangherlini solution (i.e., a
Reissner-Nordstr\"om solution in d-dimensions), one is able to give a
general definition for gravitational mass for this kind of
relativistic systems and find a mass relation with the several
quantities of the interior solution. It is also shown that for
sources of finite extent the mass is identical to the Tolman mass.


\end{abstract}
\maketitle


\section{Introduction}
\label{introd}

\subsection{Weyl type systems: Definition and overview}

Oddly, there is no noticeable mention of results for
charged fluids in Newtonian gravitation in the 19th century, an epoch
when such studies could certainly be performed with ease. Not even
the useful monograph of Ramsey of 1940 ``Newtonian attraction''
\cite{ramsey} attempts such an incursion. The first work treating
charged fluids in Newtonian gravitation we are aware of is Bonnor's
work \cite{bonnor80}, which is inspired in relativistic gravitational
works.  This work \cite{bonnor80} is divided into two distinct parts.
The first one deals with charged matter theorems in Newtonian
gravitation, and some interesting singular solutions are
displayed. The second one studies axisymmetric solutions for charged
systems in general relativity. Although the displaying of exact
solutions with matter for axisymmetric systems in general relativity
is of great interest, here we are only interested in the first part of
Bonnor's work.  In the study of Newtonian systems, Bonnor was
inspired by previous works in general relativity, indeed no
previous paper in Newtonian theory is cited.  Bonnor \cite{bonnor80}
observed that in Newtonian mechanics and classical electrostatics an
ensemble of $N$ particles of masses and charges $m_j$ and $q_j$ will
be in equilibrium in any configuration if $q_j =
\epsilon\sqrt{G\,}m_j$ ($j=1$, ..., $N$), where $\epsilon=\pm 1$ and
$G$ is the Newton's gravitational constant. In the case of continuous
distributions of charged matter, with mass density $\rho_{\rm m}$ and
charge density $\rho_{\rm e}$, there will be equilibrium everywhere if
$ \rho_{\rm e} = \epsilon\sqrt{G\,}\rho_{\rm m} $.  Such a neutral
equilibrium is possible due to the exact balancing of the
gravitational and electric forces on every fluid particle. Thus, a
static distribution of charged dust, i.e., a perfect fluid with zero
pressure, of any shape can in principle be built. Using the properties
of the Newton-Coulomb system of equations, Bonnor showed that all
Newton-Coulomb nontrivial solutions with closed equipotentials not
satisfying the relation $\rho_{\rm e}= \epsilon\sqrt{G\,}\rho_{\rm m}$
are singular.  As a byproduct he showed that a relation between the
Newtonian gravitational potential $V$ and the electric potential $\phi$,
\beq
V=V(\phi) \,,
\label{bonnoransatz}
\eeq
should have a simple form, namely,
\beq
V= -\epsilon\sqrt{G\,} \, \phi + {\rm
const}\,.
\label{bonnorrelation}
\eeq

This type of general theorems and results for the coupling between
charged matter and gravitation were originally attempted within the theory
of general relativity.  It was Weyl \cite{weyl}, while studying
electric fields in vacuum Einstein-Maxwell theory in four-dimensional
spacetime, who first perceived that it is interesting to consider a
functional relation between the metric component $g_{tt}\equiv
W^2(x^i)$ and the electric potential $\phi(x^i)$ (where $x^i$
represent the spatial coordinates, $i=1,2,3$) given by the ansatz
\beq
W=W(\phi) \,.
\label{weylansatz}
\eeq
Systems, either vacuum or matter, in which the ansatz 
(\ref{weylansatz}) holds generically will be called Weyl type systems.
If it is assumed the system is vacuum and axisymmetric 
then Weyl \cite{weyl} found that such a relation must be of the form
\beq
W^{2}= \left(-\epsilon\sqrt{G\,}\phi + b\right)^2 +c \, ,
\label{weyloriginalrelation}
\eeq
where $b$ and $c$ are arbitrary constants, and we use units such that
the speed of light equals unity.
The metric may be written as
\begin{equation}
ds^2 = - W^2 dt^2 + h_{ij}\,dx^idx^j\, ,
\quad\quad i,j=1,2,3\;, \label{metric4d}
\end{equation}
where $h_{ij}$ is also a function of the spatial coordinates $x^i$ only
and it was assumed by Weyl \cite{weyl} to represent an axisymmetric space.
Majumdar \cite{majumdar}
extended this result by showing that it holds for a large class of
static spacetimes with no particular spatial symmetry, axial or
otherwise.
Moreover, by choosing $c=0$ in which case the
potential $W^2$ assumes the form of a perfect square and so
\beq
W =-\epsilon\sqrt{G\,}\phi+ b ,\label{mppotentials4d}
\eeq
Majumdar \cite{majumdar} was able to show that the Einstein-Maxwell
equations in the presence of charged dust, i.e., a perfect fluid with
zero pressure, imply exactly the same condition of the Newtonian
theory, namely,
$
\rho_{\rm e} = \epsilon\sqrt{G\,}\rho_{\rm m} 
$, 
with both the gravitational
potential $W$ and the electric potential $\phi$ satisfying a
Poisson-like equation.  As in the Newtonian case, the relativistic
solutions are static configurations of charged dust 
and need not have any spatial symmetry.  Majumdar
\cite{majumdar} also showed that in the case $W$ is as in
Eq.~(\ref{mppotentials4d}) the metric of the three-space is conformally
flat with conformal factor given by $1/W^2$, and in
such a case all the stresses in the charged matter vanish.  Similar
results were found independently by Papapetrou \cite{papapetrou},
who assumed as a starting point a perfect square relation among $W$ and
$\phi$, in a dust filled spacetime, and showed further that the charge
density $\rho_{\rm e}$ and the relativistic energy density $\rho_{\rm m}$
are related by $\rho_{\rm e} = \epsilon\sqrt{G\,}\rho_{\rm m}$.
Relation (\ref{mppotentials4d}) is called the Majumdar-Papapetrou
relation.  Solutions in which the condition
$
\rho_{\rm e} = \epsilon\sqrt{G\,}\rho_{\rm m} 
$
and the relation (\ref{mppotentials4d}) hold are called
Majumdar-Papapetrou solutions. These kind of charged dust fluids were
studied by other authors. For instance, Das \cite{das62} showed that
if the ratio $\rho_{\rm e}/\rho_{\rm m} = \epsilon\sqrt{G\,}$ holds,
then the relation between potentials must be as in
Eq.~\eqref{mppotentials4d}.  De and Raychaudhuri \cite{deraychaudhuri}
generalized this by showing that if there is a closed equipotential
within the charged dust fluid with no singularities nor alien matter
inside it, where alien matter was an expression used to indicate
anything other than charged dust, then the charged dust fluid
corresponds to a Majumdar-Papapetrou solution.

A further advance was performed by Gautreau and Hoffman
\cite{gautreau}. They investigated the structure of the sources that
produce Weyl type fields, which satisfy the Weyl quadratic relation
\eqref{weyloriginalrelation}, in the case the matter stresses, i.e.,
the pressures, do not vanish. They found that when there is pressure,
for $W$ being given by the Weyl relation then the fluid obeys the
condition $b\,\rho_{\rm e} = \epsilon\sqrt{G\,} \left(\rho_{\rm m}+
\frac{{\rm d}-1}{{\rm d}-3}\,p \right)W +\epsilon\sqrt{G\,}
\phi\rho_{\rm e}$, or equivalently $\rho_{\rm
e}\left(\epsilon\sqrt{G\,}\phi - b\right) = -\epsilon
\sqrt{G\,}\left(\rho_{\rm m}+ \frac{{\rm d}-1}{{\rm d}-3}\,p
\right)W$, in the same spirit of the Majumdar-Papapetrou
condition. If, instead of a Weyl relation, one has a
Majumdar-Papapetrou relation, but still keeping the pressure, then the
condition is simpler, $\rho_{\rm e} = \epsilon\sqrt{G\,}\,
\left(\rho_{\rm m}+ \frac{{\rm d}-1}{{\rm d}-3}\,p \right)$.

Another interesting study was performed by Guilfoyle \cite{guilfoyle}
who considered charged fluid distributions and made the hypothesis
that the functional relation between the gravitational and the
electric potential, $W=W(\phi)$, is slightly more general than 
the one given in \eqref{weyloriginalrelation}. 
This Weyl-Guilfoyle relation has the form
\beq
W^{2}= a\left(-\epsilon \sqrt{G\,}\,\phi+ b\right)^2 +c \, ,
\label{weylguilfoyle}
\eeq
where $a$, $b$ and $c$ are arbitrary constants.  Guilfoyle
\cite{guilfoyle} investigated several general properties of such
systems.  In particular, he showed that if the fields satisfy the
Majumdar-Papapetrou perfect square relation, then the fluid pressure
is proportional to gravitational potential $p= k W^2$. In addition, a
set of exact spherically symmetric solutions were also analyzed.

Now, global analyses are important. To connect these general local
results just mentioned to general global results one needs an exterior
solution. For Weyl and Weyl-Guilfoyle relations, fluids which respect
spherical symmetry can be joined to an exterior Reissner-Nordstr\"om
solution \cite{reissnernordstrom1,reissnernordstrom2} characterized by
a mass $m$ and charge $q$.  Guilfoyle \cite{guilfoyle} by applying
junction conditions found how the global mass $m$ and the global
charge $q$ are linked to the interior fluid parameters. In these
spacetimes, whose fluids obey a Weyl-Guilfoyle relation, the total
mass $m$ is not, in general, proportional to the total electric charge
of the system $q$.  In the perfect square case of the
Majumdar-Papapetrou relation, it can be shown that the mass $m$ of the
solution is equal to its charge $q$, $\sqrt{G}m=\epsilon q$, and the
exterior is extremal Reissner-Nordstr\"om.  One can also follow the
approach of Gautreau and Hoffman \cite{gautreau}, an approach which
does not use junction conditions.  These authors have showed that, for
a source of finite extent, not necessarily spherically symmetric,
whose fluid admits pressure and a Weyl relation, and whose spacetime
is asymptotically Reissner-Nordstr\"om, the total gravitational mass
is proportional to the total electric charge of the source, and the
ratio between them is equal to the constant $b$.  Here, in this
approach, the mass definition follows from Tolman \cite{tolman}, and
can be applied to charged spacetimes (see Whittaker \cite{whittaker}
and Florides \cite{florides1,florides2}, see also \cite{cohen2}).  We
further note that the complete global understanding of a single
Majumdar-Papapetrou particle, i.e., the extremal Reissner-Nordstr\"om
vacuum solution, was achieved by Carter \cite{carter66}, through a
Carter-Penrose diagram, and the generalization to the vacuum solution
with many extremal black holes scattered around was performed by
Hartle and Hawking \cite{hawking}.

All the theorems and results mentioned above were performed in four
dimensions. It is important nowadays to have results in $\rm d$
dimensions. Indeed, string theory, the AdS/CFT conjecture, brane world
scenarios, and many other instances, point to the possible existence
of a $\rm d$-dimensional world, with $\rm d\geq4$. For charged fluids
a first attempt in the study of d-dimensional spacetimes was performed
by Lemos and Zanchin \cite{lemoszanchin5} where the beautiful
four-dimensional results of Majumdar \cite{majumdar} were put into
higher dimensions. It is thus of interest to render the general
theorems and results mentioned above into $\rm d$ dimensions.  We will
see that some of the theorems are trivially extended, but in the
process, we will get really new interesting theorems and results. For
instance, when connecting the Gautreau-Hoffman work \cite{gautreau} to
Guilfoyle's \cite{guilfoyle} we step upon a nontrivial generalization
and find an important new relation for the matter and field quantities
for charged fluids with pressure, 
which obey the Weyl-Guilfoyle relation, both in
Newtonian gravitation and general relativity. 
In this setting the notion of alien matter has to be generalized
to indicate anything other than a charged fluid of Weyl
type, guaranteeing that when the pressure is zero one recovers 
the definition in \cite{deraychaudhuri}.
We will also connect the
general local results to general global results, using as exterior
solution the d-dimensional Reissner-Nordstr\"om solution, also called
the Tangherlini solution \cite{tangherlini}. We find the d-dimensional
version of the mass relation found by Guilfoyle \cite{guilfoyle} as
well as the generalization of the Gautreau-Hoffman mass
\cite{gautreau}. Summing up, we will enounce twelve theorems and two
corollaries, and analyze some global issues.

The literature in charged fluids in general relativity is vast and we
can only mention some works more related to our paper. Inspired in the
work of De and Raychaudhuri \cite{deraychaudhuri}, some papers dealt
with scalar charge rather then Maxwell charge, see, e.g.,
\cite{wolk,teixeira,tiwari,banerjee}.  For fluids with pressure of
Majumdar-Papapetrou type it is worth mentioning here a study by Ida
\cite{ida00}. By assuming a linear equation of state for the fluid,
i.e. $\rho_ {\rm m} = {\rm const}\times p\, (= {\rm const}\times
\phi^{-2} )$, he found that the resulting equation for the electric
potential is simply a Helmholtz equation in a space of constant
curvature, and investigated some particular solutions in that case.
In four dimensions, many other works have been performed for charged
fluids with nonzero pressure without considering Weyl or other type of
relations, see, e.g., \cite{ivanov02} for a list of references.  For a
study of d-dimensional Majumdar-Papapetrou fluids and related Bonnor
stars, with a thorough review of the subject see
\cite{lemoszanchin2008}.

\subsection{Nomenclature}

We consider a ${\rm d}$-dimensional spacetime, both in Newtonian
gravitation and general relativity.  The number $n$ of space
dimensions is then $n=({\rm d}-1)$. Throughout the paper we set the speed
of light equal to unity. For the ${\rm d}$-dimensional spacetime
gravitational constant we write $G_{\rm d}$, see, e.g.,
\cite{lemoszanchin2008} for the definition of $G_{\rm d}$.  In $\rm d=4$
dimensions we write $G_4\equiv G$.

\subsubsection{Newton-Coulomb charged fluids}

Here we set out the nomenclature related to the potentials for the
Newton-Coulomb theory with charged matter. This nomenclature is
inspired in the nomenclature for the general relativity theory with
charged matter.

\begin{itemize}
\item The gravitational Newtonian potential is represented by $V$, and the
electric Coulombian potential is denoted by $\phi$. $\rho_{\rm m}$ is the
mass density, $p$ is the fluid pressure, and $\rho_{\rm e}$ is the electric
charge density.

\item $V=V(\phi)$ is, inspired in the relativistic theory, the Weyl
ansatz for the Newton-Coulomb theory.  Systems bearing this hypothesis
are Newton-Coulomb Weyl type systems, or simply Newtonian Weyl type
systems.

\item $V = -\epsilon\,\beta\,\sqrt{G_{\rm d}\,}\phi+ \gamma$ is, inspired
in the relativistic theory, what we call the original Weyl relation in the
Newton-Coulomb theory, or simply Weyl relation. The Weyl relation is, in
the Newtonian theory, the same as the Weyl-Guilfoyle relation.

\item $V = -\epsilon\, \sqrt{G_{\rm d}\,}\phi+ \gamma$ is, inspired in
the relativistic theory, what we call the Majumdar-Papapetrou relation
for the Newton-Coulomb theory (see \cite{lemoszanchin2008}). 
The Majumdar-Papapetrou relation is a particular case 
of the Weyl-Guilfoyle relation,  one sets $\beta=1$.
\end{itemize}

Here we set out the nomenclature related to the fluid quantities. For the
densities the nomenclature is more complicated than for the potentials,
since it depends on whether there is pressure or not in the fluid.

\begin{itemize}

\item When there is no pressure, the only static solutions for
$V(\phi)$ are of the form of the Majumdar-Papapetrou relation, which
for the Newton-Coulomb theory is a particular case of the
Weyl-Guilfoyle relation with $\beta=1$. In such a case, the fluid
densities obey the Majumdar-Papapetrou condition, $\rho_{\rm e} =
\epsilon\sqrt{G_{\rm d\,}}\, \rho_{\rm m}$.

\item When there is pressure, for $V$ obeying the Weyl-Guilfoyle
relation, the fluid densities obey the condition ${\beta} \rho_{\rm e}=
\epsilon \sqrt{G_{\rm d}}\, \rho_{\rm m}$, a condition given here for
the first time. Inspired in the relativistic case, we call it the
Gautreau-Hoffman condition for the Newton-Coulomb theory. Now, in the
particular case where $\beta=1$ the potential $V$ obeys the
Majumdar-Papapetrou relation for the Newton-Coulomb theory, the pressure
vanishes, and the densities are related by the Majumdar-Papapetrou
condition for the Newton-Coulomb theory.

\end{itemize}

\subsubsection{Relativistic charged fluids}

Here we set out the nomenclature related to the potentials for relativistic
fluids.

\begin{itemize}
\item In d spacetime dimensions, we use the same notation for the
metric as in four dimensions, cf. Eq.~\eqref{metric4d}, and the
symbols for electric and fluid quantities are the same as in the
Newton-Coulomb case. So, $W$ is the metric potential related to the
time coordinate, which can be interpreted as the relativistic
gravitational potential, $\phi$ is the relativistic electric
potential, i.e., the only nonzero component of the electromagnetic
gauge potential, $\rho_{\rm m}$ is the relativistic mass-energy
density, $p$ is the fluid relativistic pressure, $\rho_{\rm e}$ is the
relativistic electric charge density.

\item $W=W(\phi)$ is the Weyl ansatz. Systems bearing this hypothesis are
Weyl type systems.

\item $W^2=a\,\left(-\epsilon\sqrt{G_{\rm d}}\phi+b\right)^2 +c$
is the Weyl-Guilfoyle relation, where $a$, $b$, and $c$ are constant
parameters.   

\item $W^2= \left(-\epsilon\,\sqrt{G_{\rm d}\,}\phi+ b\right)^2 +c$ is
the original Weyl relation, or simply Weyl relation.  The Weyl
relation is a particular case of the Weyl-Guilfoyle relation.

\item $W^2= \left(-\epsilon\,\sqrt{G_{\rm d}\,}\phi+ b\right)^2$, or
$W= -\epsilon\,\sqrt{G_{\rm d}\,}\phi+ b $, is the
Majumdar-Papapetrou relation, a particular case of the Weyl relation.  

\end{itemize}

Here we set out the nomenclature related to the fluid quantities.  For the
densities the nomenclature is more complicated than for potentials,
since it depends on whether there is pressure or not in the
relativistic fluid.

\begin{itemize}

\item When there is no pressure, the only possible static solutions
satisfy the Majumdar-Papapetrou relation, and the fluid variables are
related by the equation $\,\rho_{\rm e}\left(-\epsilon\sqrt{G_{\rm
d}}\phi + b\right) =\epsilon\sqrt{G_{\rm d\,}}\,\rho_{\rm m} W$, which
can be cast into the form $ \rho_{\rm e} = \epsilon\,\sqrt{G_{\rm
d}}\, \rho_{\rm m}$. This equation between the densities is called the
Majumdar-Papapetrou condition.

\item
\begin{itemize}

\item When there is pressure, for $W$ obeying the Weyl-Guilfoyle
relation, the fluid quantities satisfy the condition $a\,b\, \rho_{\rm
e} = \epsilon\sqrt{G_{\rm d\,}}\,\left[\left(\rho_{\rm m}+ \frac{{\rm
d}-1}{{\rm d}-3}\,p \right)W + \phi\rho_{\rm e}\right]
+\epsilon\sqrt{G_{\rm d\,}}\, \left(a-1\right)\left(\phi\rho_{\rm e}
-W \rho_{\rm em}\right)$, or equivalently, $a\,\rho_{\rm
e}\left(-\epsilon\sqrt{G_{\rm d}}\phi + b\right) =
\epsilon\sqrt{G_{\rm d\,}}\, \left[\rho_{\rm m}+ \frac{{\rm d}-1}{{\rm
d}-3}\,p + \left(1-a\right)\rho_{\rm em} \right]W$, which is given
here for the first time. This condition can be considered as an
equation of state.  

\item When there is pressure, in the particular case where $a=1$, $W$
obeys the original Weyl relation and the fluid quantities are related
by $ b\, \rho_{\rm e}= \epsilon\sqrt{G_{\rm
d\,}}\,\left[\left(\rho_{\rm m}+ \frac{{\rm d}-1}{{\rm d}-3}\,p
\right)W +\phi\rho_{\rm e}\right]$, or equivalently, $\,\rho_{\rm
e}\left(-\epsilon\sqrt{G_{\rm d}}\phi + b\right) = \epsilon
\sqrt{G_{\rm d\,}}\, \left(\rho_{\rm m}+ \frac{{\rm d}-1}{{\rm
d}-3}\,p \right)W$. This is called the Gautreau-Hoffman condition.

\item When there is pressure, in the particular case where $a=1$ and
$c=0$, $W$ obeys the Majumdar-Papapetrou relation, $W =
-\epsilon\,\sqrt{G_{\rm d}}\, \phi+b$, and the fluid variables are
related by $\,\rho_{\rm e}\left(-\epsilon\sqrt{G_{\rm d}}\phi +
b\right) = \epsilon\sqrt{G_{\rm d\,}}\, \left(\rho_{\rm m}+ \frac{{\rm
d}-1}{{\rm d}-3}\,p \right)W$, or equivalently, $\rho_{\rm e} =
\epsilon\sqrt{G_{\rm d\,}}\left(\rho_{\rm m}+ \frac{{\rm d}-1}{{\rm
d}-3}\,p \right)$. This equation is a particular case of the
Gautreau-Hoffman condition. 
\end{itemize}
\end{itemize}

In \cite{deraychaudhuri} the expression alien matter was used to
indicate anything other than charged dust. Here, since we treat
charged fluids with pressure, the expression alien matter is
generalized to indicate anything other than a charged fluid of Weyl
type. When the pressure is zero one recovers the definition of
\cite{deraychaudhuri}.

The nomenclature will become clearer during the paper.

\subsection{Structure of the paper}

In Sec.~\ref{sect-newtonfluidmodel} a self-gravitating static charged
fluid in Newtonian physics is analyzed. We set up the Eulerian
formulation of fluid dynamics in $n=(\rm d-1)$-dimensional Euclidean
space.  We then analyze the general properties of a charged static
Newton-Coulomb zero pressure fluid, or charged dust, and state two theorems
rendering into higher dimensions previous results in three-dimensional
spaces by other authors. We perform an analysis of charged
Newton-Coulomb fluids with nonzero pressure reporting new results and
stating and proving four theorems.  We define the total mass and total
electric charge in terms of the respective densities and find a
mass-charge relation for Newton-Coulomb Weyl type fluids in $(\rm
d-1)$-dimensional Euclidean spaces.

In Sec.~\ref{sect-relatfluidmodel} a self-gravitating static charged
fluid in general relativity is analyzed.  The general formalism and
set up is presented, followed by a study of a zero pressure charged fluid,
or charged dust, where two theorems are stated and proved, rendering into
higher dimensions results by Majumdar \cite{majumdar}, Papapetrou
\cite{papapetrou}, Das \cite{das62}, and De and Raychaudhuri
\cite{deraychaudhuri}.  The study of a relativistic charged pressure
fluid of Weyl type is divided into six parts.  In the first part the
basic equations are written and general results are reported.  A
theorem rendering into higher dimensions some results obtained in $\rm
d=4$ by Guilfoyle \cite{guilfoyle} is stated there. In the second part
the Weyl ansatz is imposed and its consequences to the relativistic
fluid are analyzed. Another theorem rendering into higher dimensions
some other results obtained in $\rm d=4$ by Guilfoyle \cite{guilfoyle}
is stated in this part.  For a Weyl-Guilfoyle relation, a generalized
condition obeyed by the mass density, pressure, electric charge
density, and electromagnetic energy density of a fluid in four and
higher dimensions is found.  This is done in the third part where a
new theorem is proved. The fourth part contains the same new theorem
on Weyl type fluids in $\rm d$-dimensional spacetimes, the proof being
now inspired in the work by De and Raychaudhuri
\cite{deraychaudhuri}. In the fifth part we study spherically
symmetric Weyl type systems and compare with the work in four
dimensions done by Guilfoyle \cite{guilfoyle}. In the sixth part an
analysis is performed where we show that the relativistic Weyl type
charged pressure fluids have a correct Newtonian limit consistent with
what was found in Sec. \ref{sect-newtonfluidmodel}. Then closing
Sec.~\ref{sect-relatfluidmodel}, the mass and charge definitions for
relativistic Weyl type fluids in asymptotically Tangherlini spacetimes
are given, and the particular case of spherical symmetry is studied in
some detail.

In Sec.~\ref{conclusions} we finally state our conclusions.

\section{Newton-Coulomb charged fluid with pressure in ${\rm d}=n+1$
spacetime dimensions}
\label{sect-newtonfluidmodel}

\subsection{Basic equations}
\label{sect-basic_newton}

Consider a ${\rm d}$-dimensional Newtonian spacetime,  with 
the number of space dimensions  $n$ being $n={\rm d}-1$.
Let us study the dynamics of a gravitating Newtonian charged pressure
fluid in a $({\rm d}-1)$-dimensional Euclidean space according to the Euler
description. The basic equations are the continuity and the Euler
equations, which may be written as
\beqa
&& \frac{\partial \rho_{\rm m}}{\partial t} +
\nabla_i\left(\rho_{\rm m}\,
v^i\right)=0 \, , \label{continuity}\\
&&\rho_{\rm m}\frac{\partial v_i}{\partial t} +\rho_{\rm m} v^j\,\nabla_j
 v^i +\nabla_i p = -\rho_{\rm m}\nabla_i V -
\rho_{\rm e}\nabla_i \phi
\, ,\label{euler}
\eeqa
where $t$ is the time coordinate, $\nabla_i$ is the $({\rm
d}-1)$-dimensional gradient operator. The fluid quantities appearing in the
above equations are, the $({\rm d}-1)$-dimensional fluid velocity $v_i$,
the matter density $\rho_{\rm m}$, the electric charge density $\rho_{\rm
e}$, and the fluid pressure $p$. Finally,  $V$ and $\phi$ are respectively
the Newton gravitational and the Coulomb electric potentials, given by the
respective Poisson equation
\beqa
&&\nabla^2 V = S_{\rm d-2}\,G_{\rm d}\,\rho_{\rm m}\, ,
\label{gravitypoisson}\\
&&\nabla^2 \phi = - S_{\rm d-2}\, \rho_{\rm e}\, , \label{electricpoisson}
\eeqa
where the operator $\nabla^2$ is the Laplace operator in $({\rm
d}-1)$-dimensional Euclidean space, $S_{\rm d-2}= 2\pi^{({\rm
d}-1)/2}/\Gamma(({\rm d}-1)/2)$ is the area of the unit sphere
$\mathbf{S}^{{\rm d}-2}$, $\Gamma$ is the usual gamma function, and
$G_{\rm d}$ is the Newton's gravitational constant in ${\rm d}$
dimensions.  $S_{\rm d-2}$ reduces to $4\pi$ in three space (four
spacetime) dimensions, and Eqs.~(\ref{gravitypoisson}) and
(\ref{electricpoisson}) are the natural generalizations of the
corresponding three-dimensional Poisson equations for the potentials
$V$ and $\phi$ to $({\rm d}-1)$-dimensional space.

We will consider only static systems, so all of the quantities are
functions of the space coordinates only, and the fluid's velocity can
be made equal to zero, $v_i=0$. Then, the continuity equation is
identically satisfied and the Euler equation for the charged pressure
fluid in static equilibrium reads
\beq
\rho_{\rm m} \,
\nabla_i\, V
+ \rho_{\rm e}\,\nabla_i\,\phi+\nabla_i\, p
 =0\, .
\label{eulerstatic}
\eeq
The important equations for the problem are
Eqs.~(\ref{gravitypoisson})-(\ref{eulerstatic}). 
Eqs.~\eqref{gravitypoisson} and
\eqref{electricpoisson} are the field equations that determine the
gravitational and the electric potentials once the mass and charge densities
are given, while Eq.~(\ref{eulerstatic}) is the equilibrium
equation for the system. Some particular cases of these equations are
considered below.

\subsection{Zero pressure: Weyl type systems in the Newton-Coulomb theory
and the Bonnor theorem in higher dimensions}
\label{sect-bonnordd}

A special but very interesting case of charged matter is the dust fluid,
for which $p=0$, where Eq.~\eqref{conserveq} reduces to
\beq
\nabla_i\, V +\frac{\rho_{\rm e}}{\rho_{\rm m}} \,\nabla_i\,\phi\
=0\, .\label{eulereqp0}
\eeq 
From Eq.~\eqref{eulereqp0} we can render into $\rm  (d-1)$-dimensional
spaces a theorem by Bonnor in $\rm d=4$. Bonnor himself was inspired in the
relativistic analysis by De and Raychaudhuri \cite{deraychaudhuri}.

\begin{thm} {(Bonnor 1980)}
\label{bonnor0}

\noindent
For any charged static dust distribution in the Newton-Coulomb theory,
the $({\rm d}-2)-$dimensional hypersurfaces of constant $V$ coincide with
the $({\rm d}-2)-$dimensional hypersurfaces of constant $\phi$, and $V$
is a function of $\phi$ alone.
\end{thm}

\begin{proof}
Note that $V$ and $\phi$ are scalar functions in a $({\rm
d}-1)$-dimensional Euclidean space, so any one of the conditions of
constant $V$ or $\phi$ defines a $({\rm d}-2)$-dimensional space. Now
contracting Eq.~\eqref{eulereqp0} with $dx^i$ it gives $\rho_{\rm m}
\,(\nabla_i\,V) dx^i + \rho_{\rm e}\,(\nabla_i\,\phi) dx^i =0$, which means
$\rho_{\rm m} \,dV + \rho_{\rm e}\, d\phi =0$. This equation implies that
if any one of $dV$ or $d\phi$ is null, then both of them are. In other
words, it results $dV/d\phi = -\rho_{\rm e
}/\rho_{\rm m}$, which means that $dV/d\phi= \partial V/\partial\phi$.
\end{proof}

\noindent
This results means that for a Newton-Coulomb charged dust fluid in
which $p=0$ there is a functional relation between the gravitational
potential $V$ and the electric potential $\phi$:
\beq
V=V(\phi)\, . \label{weylhyp}
\eeq
Eq.~(\ref{weylhyp}) is the Weyl ansatz which tells also that $\phi$
and $V$ share the same equipotential surfaces.  Theorem
\eqref{bonnor0} is the $(\rm d-1)$-dimensional version a theorem by
Bonnor
\cite{bonnor80} in three-dimensional space. The theorem by Bonnor
\cite{bonnor80} is the Newtonian version of a theorem stated by De and
Raychaudhuri for four-dimensional relativistic charged dust fluids
\cite{deraychaudhuri} (see Sec.~\ref{sect-deraydd}). The 
inclusion of pressure for relativistic fluids was done by Guilfoyle
\cite{guilfoyle} (see Sec.~\ref{sect-nonzerop-relat1}).

We then work out the basic equations for the dust fluid case using the
general form \eqref{weylhyp}. To begin with, let us rewrite 
Eq.~\eqref{gravitypoisson} by taking the Weyl ansatz~\eqref{weylhyp}
into account. It assumes the form
\beq
{V}^\prime \,\nabla^2\,\phi+ V''\, \left(\nabla_i\,\phi\right)^2 =
S_{\rm d-2}G_{\rm d}\,\rho_{\rm m}\, ,\label{gravityfunct}
\eeq
where the primes stand for derivatives with respect to $\phi$.
Of course, Eq.~\eqref{electricpoisson} does not change form. On the other
hand, the equilibrium equation \eqref{eulerstatic}  now reads
\beq
\rho_{\rm m} \, V^\prime+ \rho_{\rm e}
=0\, .\label{eulerfunctp0}
\eeq
Substituting $\rho_{\rm m}$ from the last equation,
Eq.~(\ref{eulerfunctp0}), into Eq.~\eqref{gravityfunct}, we find
$
\left({V}^\prime\right)^2 \,\nabla^2\,\phi+
V'\,V''\,
\left(\nabla_i\,\phi\right)^2 =
-S_{\rm d-2}G_{\rm d}\,\rho_{\rm e}\, .$
Then with the help of Eq.~\eqref{electricpoisson} we find
\beq
\left[V^\prime\,^2-\,\,G_{\rm d}\right]\nabla^2\,\phi+
V'\,V''\,\left(\nabla_i\phi\right)^2=0\, .\label{eulerfunct3p0}
\eeq
This equation can be cast into the form
\beq
Z\,\nabla_i\left(Z\, \nabla^i\phi\right)  =0 ,
\label{bonnortheop0eq}
\eeq
where we have assumed $V^\prime\,^2>\,\,G_{\rm d}$ and defined
\beq
Z= \sqrt{V^\prime\,^2-\,\,G_{\rm d}\,} \, ,\label{Znewtonp0}
\eeq
and, otherwise, if $V^\prime\,^2-G_{\rm d} < 0$, one just needs to redefine
$Z$ as $Z= G_{\rm d}-V^\prime\,^2$.
Now we can state a theorem rendering into higher dimensions a theorem by
Bonnor \cite{bonnor80} in tree-dimensional space. The analysis by Bonnor
is inspired in the relativistic analysis of Papapetrou \cite{papapetrou},
Das \cite{das62}, and
De and Raychaudhuri \cite{deraychaudhuri}.

\begin{thm} {(Bonnor 1980)}
\label{theoNewt1}

\noindent (i) In the Newton-Coulomb theory,
if the surfaces of any static charged dust distribution are
closed equipotential hypersurfaces and inside these hypersurfaces there are
no singularities, holes or alien matter, then the function $V(\phi)$ must
satisfy the relation
\beq
 V = - \epsilon\sqrt{G_{\rm d}}\, \phi\, +\gamma  ,\label{mprelationp0}
\eeq
with $\gamma$ being an integration constant, and in such a case it 
follows that
\beqa
& &\rho_{\rm e} = \epsilon\sqrt{G_{\rm d}}\, \rho_{\rm
m}\,.
\label{mpconditionp0}
\eeqa

\noindent {(ii)}
In the Newton-Coulomb theory,
if the ratio $\rho_{\rm e}/\rho_{\rm m}$ equals a constant $K$, and there
are no singularities, holes, nor alien matter in that region,
then it follows the relation \eqref{mprelationp0} for the potentials, and
also $K= \epsilon\sqrt{G_{\rm d}}$.
\end{thm}

\begin{proof}
The proof of {\it (i) } is as follows: Assuming $Z\neq 0$,  one
obtains $\nabla_i\left(Z\nabla^i\phi\right)=0$. Then define a new field
$\psi$ by $\nabla^i\psi = Z\nabla^i\phi$, which due to 
Eq.~\eqref{bonnortheop0eq} is divergenceless, i.e., $\nabla^2\psi=0$. Now
integrate the quantity $\nabla_i\left(\psi\nabla^i\psi\right)$ over a
finite volume ${\cal V_S}$ in $({\rm d}-1)$-dimensional  space to get
\beqa
\int_{\cal V_S}\nabla_i\left(\psi\nabla^i\psi\right)\,d{\cal V} =
\int_{\cal V_S}\left(\nabla_i\psi\right)^2\,d{\cal V} =
\int_{\cal S} \psi\,\left(\nabla_i\psi\right)\, n^i d\cal S\, ,
\label{teop0proof1}
\eeqa
${\cal S}$ being the boundary of ${\cal V_S}$, $n^i$ is the unit vector
normal to $\cal S$, and we have used the Gauss theorem. If there exist a
closed surface which is an equipotential surface for $\phi$, then by
identifying such a surface with $\cal S$ one finds
\beq
\int_{\cal V_S}\left(\nabla_i\psi\right)^2\,d{\cal V} =
\int_{\cal V_S} Z^2 \left(\nabla_i\phi\right)^2\,d{\cal V} =
 0\, . \hfill \label{theop0proof2}
\eeq
Since the integrand is a positive definite function in which
$\nabla_i\phi\neq0$, the integral in Eq.~\eqref{theop0proof2} gives
zero only if $Z=0$ everywhere within the region bounded by $\cal
S$. The condition $Z=0$ implies in $V'= -\epsilon\sqrt{G_{\rm d}}$,
with $\epsilon=\pm 1$. This result substituted into
Eq.~\eqref{eulerfunctp0} furnishes Eq.~\eqref{mpconditionp0} and
after integration it gives Eq.~\eqref{mprelationp0}, completing the proof
of {\it (i)}.  The given proof follows standard proofs in 
potential theory where 
the uniqueness of the solutions of Poisson equation under Dirichlet
or Neumann boundary conditions is discussed.
The proof of {\it (ii)} is as follows: Assuming the ratio $\rho_{\rm
e}/ \rho_{\rm m}$ is constant, then Eq.~\eqref{eulerfunctp0} implies
$V'=$ constant and from \eqref{bonnortheop0eq} one gets $Z\,\nabla^2
\phi =0$.  Since, in view of Eq.~\eqref{electricpoisson}, $\nabla^2
\phi \neq 0$, this leads to $Z=0$, or $V'^2 = G_{\rm d}$, resulting
the same relations among $V$ and $\phi$, and among $\rho_{\rm m}$ and
$\rho_{\rm m}$ as in Eqs.~\eqref{mprelationp0} and \eqref{mpconditionp0} 
above.
\end{proof}

\noindent
The results just presented generalize the theorem by Bonnor \cite{bonnor80}
to $({\rm d}-1)$-dimensional spaces, which is the Newtonian version of the
results found in part by Papapetrou \cite{papapetrou},
in part by Das \cite{das62}, and fully by De and Raychaudhuri
\cite{deraychaudhuri} (see Sec.~\ref{sect-deraydd}).
Eqs.~\eqref{mprelationp0} and \eqref{mpconditionp0} are called the
Majumdar-Papapetrou relation and the Majumdar-Papapetrou condition,
respectively.

\subsection{Nonzero pressure: Weyl type systems in the Newton-Coulomb theory
and new theorems in higher dimensions}
\label{sect-nonzerop-newton}

\subsubsection{Static charged pressure fluid: General properties}
\label{guilfoyle}

Equation~\eqref{eulerstatic} can be written in terms of total derivatives.
For contracting it with $dx^i$ yields $\rho_{\rm m} \,(\nabla_i\, V) dx^i +
\rho_{\rm e}\,(\nabla_i\,\phi) dx^i +( \nabla_i\, p) dx^i =0$, or
$\rho_{\rm m} \,dV + \rho_{\rm e}\, d\phi+ dp=0$. This implies that $V$,
$\phi$ and $p$ are functionally related, viz, $p=p(V,\phi)$, with 
$\rho_{\rm m} = -\left(\partial p /\partial V\right)_\phi$ and $\rho_{\rm
e} =- \left(\partial p /\partial \phi\right)_V$.
With this, we can state a theorem, whose relativistic version in four
dimensions can be found in Guilfoyle \cite{guilfoyle}.

\begin{thm} (Newtonian version of Guilfoyle 1999)
\label{theo-guilf261}
\item
 For any static charged pressure fluid in the Newton-Coulomb theory, if any two
of the $({\rm d}-2)-$dimensional hypersurfaces of constant
$V$, $\phi$, or $p$ coincide, then the third also coincides.
\end{thm}
\begin{proof}
First observe that $V$, $\phi$ and $p$ are scalar functions in a $({\rm
d}-1)$-dimensional Euclidean space, so any one of the conditions of
constant $V$, $\phi$, or $p$ defines a $({\rm d}-2)$-dimensional space.
Moreover, from the continuity equation one gets $\rho_{\rm m} \,dV +
\rho_{\rm e}\, d\phi+ dp=0$. This equation implies that if any two of $dV$,
$d\phi$, or $dp$ are simultaneously null, then all of them are null.
\end{proof}
\noindent
Theorem \ref{theo-guilf261} generalizes theorem \ref{bonnor0}, since one is
now including pressure. For $\rm d=4$ it corresponds to the Newtonian
version of a theorem by Guilfoyle on relativistic charged pressure fluids
of Weyl type (see Sec.~\ref{sect-nonzerop-relat}). Further consequences of
this theorem are explored next.

\subsubsection{Weyl and Majumdar-Papapetrou relations for generic pressure
in the Newton-Coulomb theory in higher dimensions}
\label{weylansatz_section}

Doing for $\rm d$-dimensional Newtonian gravitation what Weyl did for
general relativity \cite{weyl} even in the presence of matter with
pressure, assume a functional relation between the gravitational and the
electric potential, as in Eq.~\eqref{weylhyp}, $V=V(\phi)$. With this 
ansatz, that we call here the Weyl ansatz, Weyl originally worked out the
Einstein-Maxwell vacuum equations that would follow and found that the
relativistic potential is a quadratic function of the electric potential.
Doing the same here in Newtonian gravitation, we find that the ansatz
\eqref{weylhyp}, when substituted into Eqs.~\eqref{electricpoisson} and
\eqref{gravitypoisson}, gives in vacuum, $\rho_{\rm m}=0$, $p=0$, and
$\rho_{\rm e}=0$, that $,V''=0$, i.e.,
\beq
V(\phi) = -\epsilon\,\beta\, \sqrt{G_{\rm d}\,}\phi + \gamma
\label{weylrelationNCvacuum}\, .
\eeq
where $\beta$ and $\gamma$ are arbitrary constants. 
This is the Weyl
relation for the Newton-Coulomb theory in vacuum. When $\beta=1$ 
one has 
$
V(\phi) = -\epsilon\, \sqrt{G_{\rm d}\,}\phi + \gamma
\label{mprelationNCvacuum}\, ,
$
which is the Majumdar-Papapetrou relation, 
see Eq.~(\ref{mprelationp0}). In 
\cite{lemoszanchin2008} it was stated that the Weyl
and the Majumdar-Papapetrou relations are the same in 
the Newton-Coulomb theory, but 
this is only true for $\beta=1$. In fact, $\beta\neq1$ 
is important when one considers systems with pressure.

Let us now work out the basic equations for matter with nonzero pressure
using the Weyl ansatz~\eqref{weylhyp}. It is
convenient to analyze first Eq.~\eqref{eulerstatic}, which now reads
$
\left(\rho_{\rm m}\,V^\prime +\rho_{\rm e}\right)
\,\nabla_i\,\phi+\nabla_i\, p
=0\, .
$
It follows that $p$ is also a function of $\phi$, $p=p(\phi)$. So, the
two fields $V$ and $\phi$ have the same equipotential surfaces, which
are also surfaces of constant pressure. Since we consider
$\nabla_i\,\phi\neq0$, Eq.~\eqref{eulerstatic} is then equivalent to
\beq
\rho_{\rm m} \, V^\prime+ \rho_{\rm e}+ p^\prime
=0\,.\label{eulerfunct}
\eeq
where again the prime stands for the derivative with respect to $\phi$.
Let us now state these results in a compact form.

\begin{thm}  (Newtonian version of Guilfoyle 1999)
\label{theo-guilf262}

\item (i) If the Newton-Coulomb charged pressure fluid is of Weyl type
and is in equilibrium, then the equipotentials are also hypersurfaces of
constant pressure, and vice-versa. 
\item (ii) If the Newton-Coulomb charged pressure fluid is of Weyl type and
is in equilibrium, then either the pressure gradient vanishes at the
surface of the fluid or the surface is an equipotential.
\end{thm}

\begin{proof} Using the Weyl ansatz $V= V(\phi)$ and the
equilibrium equation~\eqref{eulerstatic} we have  $\left(\rho_{\rm
m}\,V^\prime +\rho_{\rm e}\right) d\,\phi+d p=0$. Then the hypersurfaces of
constant $\phi$ and $V$ coincide, and theorem \ref{theo-guilf261} implies
the third, of constant $p$, also coincides. On the other hand, a surface of
constant $p$ implies $d\phi=0$, and thus it is also a surface of constant
$\phi$ and of constant $V$. This proves {\it (i)}. To prove {\it (ii)} we
note that continuity conditions establish that the pressure is zero at the
surface of the fluid, which means it is a surface of constant pressure,
$dp= (\nabla_i\, p)dx^i=0$. Then, unless the pressure gradient vanishes at
the surface, $\nabla_i p= 0$ for all $i$, by {\it (i)} the surface of the
fluid is an equipotential surface.
\end{proof}

\noindent
Guilfoyle \cite{guilfoyle} has shown an analogous theorem for relativistic
charged pressure fluid of Weyl type in four-dimensional spacetimes. The
above theorem \ref{theo-guilf262} is the Newtonian version of Guilfoyle's
theorem extended to higher dimensional spaces (see
Sec.~\ref{sect-nonzerop-relat2}).

Now we rewrite the basic equation on basis of theorems \ref{theo-guilf261}
and \ref{theo-guilf262}. From Eqs.~\eqref{electricpoisson} and
\eqref{gravitypoisson}, together with \eqref{eulerfunct}, we can find an
equation for $\phi$ in terms of $V'$ and $p'$:
\beq
\left[\left({V}^\prime\right)^2 -G_{\rm d}\right]\nabla^2\,\phi+
V'\,V''\, \left(\nabla_i\,\phi\right)^2 = -S_{\rm d-2}\,G_{\rm d}\, p'\,
. \hfill
\label{eqpprime}
\eeq
Once $V(\phi)$ and $p(\phi)$ are supplied, this is the final equation to be
solved for $\phi$. Such an analysis, however, is not done in the present
work.

\subsubsection{New theorem in four and higher dimensions and the
Weyl-Guilfoyle relation: following Gautreau-Hoffman}
\label{sect-gautreau-newton}

We turn once again to Eqs.~\eqref{gravitypoisson} and
\eqref{electricpoisson} and observe that multiplying the second of
those equations by $ -\epsilon\,\beta \sqrt{G_{\rm d}}$, with $\epsilon
= \pm 1$ and $\beta $ being an arbitrary parameter, and subtracting
the result form the first equation we get
\beq
\nabla^2\left(V + \epsilon\,\beta\sqrt{G_{\rm d}}\, \phi-\gamma \right) =
S_{\rm d-2}\sqrt{G_{\rm d}}\left( \sqrt{G_{\rm d}}\,\rho_{\rm m} -\epsilon\,
\beta\, \rho_{\rm e}\right)\, , \label{GHnewton1}
\eeq 
where $\gamma$ is a constant. Now, if $V+\epsilon\,\beta\sqrt{G_{\rm d}}\,
\phi -\gamma = 0$ everywhere inside matter, it follows that 
$\rho_{\rm e}$ is proportional to $\rho_{\rm m}$, and, conversely,
assuming the right-hand side of Eq.~\eqref{GHnewton1} is zero, i.e., if
$\rho_{\rm e}$ is proportional to $\rho_{\rm m}$, then $\nabla^2\left(V
+\epsilon\,\beta\sqrt{G_{\rm d}}\, \phi\right)=0$. Thus one can state the
following theorem:

\begin{thm} {(New)} \label{theoGH1}

\noindent
(i) In a Newton-Coulomb charged pressure fluid, if the potentials are
such that $V +\epsilon\,\beta\sqrt{G_{\rm d}} \, \phi-\gamma=0$, with
constant $\beta$ and $\gamma$, then it follows the condition
\beq
\beta\,\rho_{\rm e}= 
\epsilon {\sqrt{G_{\rm d}}}\, \rho_{\rm m}\,.
\label{chargerel1}
\eeq

\noindent
(ii) In a Newton-Coulomb charged pressure fluid, if the ratio
$\rho_{\rm e}/\rho_{\rm m}$ equals a constant and there is a closed
surface, with no singularities nor alien matter inside it, 
where $V
+\epsilon\,\beta\sqrt{G_{\rm d}}\,\phi - \gamma $ vanishes,
then it follows
\beq
V =-\epsilon\,\beta\,\sqrt{G_{\rm d}}\,\phi +\gamma
\label{mprelationGH1}
\eeq
everywhere.
\end{thm}

\begin{proof}
The proof of {\it (i)} follows straightforwardly from
Eq.~\eqref{GHnewton1},
because the hypothesis of the theorem implies that the right-hand side of
such an equation must be zero. For {\it (ii)}, we take 
$\rho_{\rm e}/\rho_{\rm m}= \epsilon \sqrt{G_{\rm d}}/\beta$, with
constant $\beta$, and then Eq.~\eqref{GHnewton1} implies in
$\nabla^2\left(V+ \epsilon\,\beta\sqrt{G_{\rm d}}\, \phi-\gamma \right)=0$.
Let $F=V+ \epsilon\,\beta \sqrt{G_{\rm d}}\, \phi -\gamma$, so that
$\nabla^2F=0$. Hence, in view of this condition one has
$\nabla_i\left(F\nabla^iF\right)= \left(\nabla_i
F\right)^ 2$. Integrate this equation over a volume ${\cal V_S}$ in
$({\rm d}-1)$-dimensional space to get
\beqa
\int_{\cal V_S}\nabla_i\left(F\nabla^i F\right)\,d{\cal V} =
\int_{\cal V_S}\left(\nabla_i F\right)^2\,d {\cal V } =
\int_{\cal S} F\,\left(\nabla_i F \right)\,n^i d{\cal S}\, ,
\label{eqtheoGH1}
\eeqa
$\cal S$ being the boundary of ${\cal V_S}$, $n^i$ being the unit vector
normal to $\cal S$, and the Gauss theorem has been used. If there exist a
closed surface on which $F$ vanishes, then by identifying such a surface
with $\cal S$ one finds $\displaystyle{\int_{\cal V_S}\left(\nabla_i
F\right)^2\,d {\cal V} =  0 }$, which is satisfied only if $\nabla_ i F=0$
everywhere inside $\cal S$. This means $F= {\rm constant}$ everywhere in
the region bounded by $\cal S$ and the stated result follows.
\end{proof}

\noindent
Our theorem \ref{theoGH1} is a Newtonian version of the results we
find in the relativistic section (see
Sec.~\ref{sec-gautreau-rel}). Our relativistic theorem for fluids
obeying a Weyl-Guilfoyle relation, in turn, was inspired on the
analysis by Gautreau and Hoffman \cite{gautreau}, who studied
relativistic charged pressure fluids obeying a Weyl relation in
four-dimensional spacetimes.  Equation~\eqref{chargerel1} may be
thought as the most general condition relating the densities and the
pressure in the Newton-Coulomb theory with matter for fluids obeying a
Weyl type relation.  We call it the Gautreau-Hoffman condition for the
Newtonian theory. Note also that Eq.~(\ref{mprelationGH1}) is
identical to the Weyl relation for the Newton-Coulomb theory in
vacuum, given by Eq.~\eqref{weylrelationNCvacuum}. However, upon
comparison with the relativistic case, it is found that it is also a
Weyl-Guilfoyle relation, so, in the case of pressure, it can be said
that the Weyl and the Weyl-Guilfoyle relations coincide in the
Newtonian case. Unlike the relativistic case, the most general
relation between potentials in the Newton-Coulomb theory, coincides
with the relation obtained in vacuum.  When one puts $\beta=1$ in
Eqs.~\eqref{chargerel1}-\eqref{mprelationGH1} one gets the
Majumdar-Papapetrou condition and relation, respectively.  As we will
see $\beta=1$ means no pressure, while arbitrary $\beta$,
$\beta\neq1$, implies fluids with nonzero pressure.

\subsubsection{The same new theorem as last subsection:
following Bonnor-De-Raychaudhuri}
\label{sect-deraynewton}

We now proof in a different way the theorem stated in the last
subsection. Here we follow Bonnor \cite{bonnor80}, who in turn
followed the relativistic theorem of De and Raychaudhuri
\cite{deraychaudhuri}.  Indeed, the result found above can also be
obtained following the Bonnor approach \cite{bonnor80}, in which the
equilibrium equation is used as a subsidiary condition. As we shall
see below, this strategy brings the pressure into play and, after some
hypotheses concerning the relation among the pressure gradient and
mass and charge densities, a result that is similar to theorem
\ref{theoGH1} is found.

Notice that Eq.~\eqref{eulerfunct} displays explicitly an analogy
between the pressure gradient $p'$ and the charge density $\rho_{\rm e}$,
in fact $\rho_{\rm e}$ can be considered as some sort of pressure
gradient, both act as to balance the gravitational attraction.  Such a
similarity becomes even more striking by substituting $\rho_{\rm m}$
from the last equation, Eq.~(\ref{eulerfunct}), into
Eq.~\eqref{gravitypoisson}, to find $\nabla^2
V=\left({V}^\prime\right)^2 \,\nabla^2\,\phi+ V'\,V''\,
\left(\nabla_i\,\phi\right)^2 = -S_{\rm d-2}\,\left( \rho_{\rm e}+p'
\right)$.  Thus, the derivative of the pressure with respect to the
electric potential $\phi$ acts as a source for the electric field, in
much the same way as the charge density does. It is then natural to
assume a relation in the form $p'= \chi\,\rho_{\rm e}$. However, since
the Weyl ansatz \eqref{weylhyp} tell us that the potentials $V$ and
$\phi$ are related to each other, so are the charge and mass densities.
It is then reasonable to try a more general relation among $p'$, $\rho_{\rm
e}$ and $\rho_{\rm m}$. 
Namely,
\begin{equation}
p'= \chi\,\rho_{\rm e} + \lambda V'\rho_{\rm m}\,, \label{pgradient}
\end{equation}
$\chi$ and $\lambda$ being arbitrary functions of the coordinates. After
this definition, using Eqs.~\eqref{electricpoisson}, \eqref{gravityfunct}
and \eqref{eqpprime}  one finds
$
\left[\left(1+\lambda\right)V^\prime\,^2-\,\left({1+\chi}\right)
\,G_{\rm d}\right]
\nabla^2\, \phi+ \left(1+\lambda\right)V'\,
V''\,\left(\nabla_i\phi\right)^2=0$ .
This equation can be cast into the form
\beq
Z_p\nabla_i\left(Z_p \nabla^i\phi\right)  =
-\frac{1}{2}\left({V^\prime}^2\lambda^\prime -G_{\rm d}\chi^\prime\right)
\left(\nabla_i\phi\right)^2\, ,\label{eqforchi}
\eeq
where we have defined
\beq
Z_p= \sqrt{\left(1+\lambda\right)V^\prime\,^2-
\,\left({1+\chi}\right)\,G_{\rm d}} \, .\label{Znewtonp_chi}
\eeq
Since a complete analysis of the general solutions to
Eq.~\eqref{eqforchi} for arbitrary $\lambda$ and $\chi$ is not an easy
task, here we will study some particular cases and leave the general
case to be considered in future work.

Consider thus the particular case of a charged pressure fluid satisfying
\eqref{pgradient} with constant $\chi$ and $\lambda$. This condition implies
$\chi'=0$ and $\lambda'=0$, and from Eqs.~\eqref{eulerfunct} and
\eqref{eqforchi} we get
\beqa
&&V'= -\frac{\left(1+\chi\right)\rho_{\rm e}} {\left(1+\lambda\right)
 \rho_{\rm m}}\, , \label{mpconditionp_chi}\\
&&Z_p \,\nabla_i \left( Z_p\,  \nabla^i\phi\right)  = 0\,.
 \label{bonnortheop_chi}
\eeqa
From this we can prove statements that generalizes theorem
\ref{theoNewt1} to include pressure,
and whose results are equivalent to what was found in
connection to theorem \ref{theoGH1}.

\begin{thm} (Same as theorem \ref{theoGH1}, following other path)
\label{theoNewt2}
 
\noindent
(i) If the surfaces of any Newton-Coulomb charged pressure fluid
distribution are closed equipotential hypersurfaces and inside these
hypersurfaces there are no holes, singularities or alien matter, and the
fluid is of Weyl type, whose pressure satisfies the condition given in
Eq.~\eqref{pgradient} with constant $\lambda$ and $\chi$, then the
relations
\beqa
& &V=-\epsilon\sqrt{\frac{1+\chi}{1+\lambda}\,G_{\rm d}\,}\,\phi
+\gamma\, , \label{Vp_chi}\\
& & \rho_{\rm e}= \epsilon\sqrt{\frac{1+\lambda}{1+\chi}G_{\rm d}\,}\,
\rho_{\rm m}\, , \label{rhop_chi}
\eeqa
with constant $\lambda$, $\gamma$ and $\chi$, hold everywhere
inside the fluid distributions.

\noindent
(ii) In a Newton-Coulomb charged pressure fluid, if the relation
between mass and charge densities is as in Eq.~\eqref{rhop_chi}, with
constant $\lambda$ and $\chi$, and there is no singularities nor alien
matter in the considered region, then the relation \eqref{Vp_chi} holds.
\end{thm}

\begin{proof}
First we show that, with the above conditions, $Z_p$ must be zero. This is
done by following the same reasoning as in the case of Theorem
\ref{theoNewt1}, where now the new field $\psi$ is defined by
$\nabla_i\psi= Z_p\nabla_i\phi$, with $Z_p$ given by Eq.
\eqref{Znewtonp_chi}. Once it is proved that $Z_p=0$, which means
$(1+\lambda){V'}^2= G_{\rm d}(1+\chi)$,
Eq.~\eqref{Vp_chi} is obtained by direct integration of this
result, while Eq.~\eqref{rhop_chi} follows from this and
from Eq.~\eqref{mpconditionp_chi}. This proves {\it (i)}.
The proof of {\it (ii)} is equivalent to Theorem \ref{theoNewt1}{\it
(ii)} and we do not give it here.
\end{proof}

\noindent
This theorem \ref{theoNewt2} is inspired in the analysis of a
relativistic charged dust fluid in four-dimensional spacetimes by
De and Raychaudhuri \cite{deraychaudhuri} (see Sec.~\ref{sect-deraydd}).
It generalizes theorem \ref{theoNewt1} to include pressure, which in
turn generalizes results by Bonnor \cite{bonnor80} for a charged
pressureless Newton-Coulomb fluid in $\rm d=4$.

As seen above, in order to have nonsingular solutions of Weyl type in
$({\rm d}-1)$-dimensional Euclidean spaces filled by charged matter
with pressure satisfying the condition $dp/d\phi = \chi \rho_{\rm
e}+\lambda\, V'\rho_{\rm m}$, with constant $\chi$ and $\lambda$, the
function $V(\phi)$ must be a linear function of $\phi$. Moreover, the
ratio between mass and charge densities is a constant. As expected
from Theorem \ref{theoNewt1}, the case $\chi=\lambda$ gives $\rho_{\rm
e}=\sqrt{G_{\rm d}\,}\, \rho_{\rm m}$ and the result is the dust fluid
$p=0$. These results are essentially the same as what was found in
connection with theorem \ref{theoGH1}, but now we have a two parameter
solution, one of them connecting the pressure to the charge density
and the other one relating the pressure to the mass density. To see
that explicitly define $\beta =\sqrt{\frac{1+\chi}{1+\lambda}\,}$ to
find that the relations \eqref{Vp_chi} and \eqref{rhop_chi} reproduce,
respectively, Eqs.~\eqref{mprelationGH1} and \eqref{chargerel1}, the
corresponding results of Theorem \ref{theoGH1}.  The two parameters
$\chi$ and $\lambda$ are convenient for the comparison of the present
results with the relativistic case.

A consequence of the above analysis is that the pressure gradient
$p^\prime$ is proportional to the charge density (or to the mass
density). Moreover, if the fluid satisfies the Majumdar-Papapetrou
relation for the potentials, the pressure is zero.  This can be stated
as a corollary to Theorems \ref{theoGH1} and \ref{theo-deray2}.

\begin{cor} (New)\label{cor-newt}

\noindent
For any static charged pressure fluid distribution in the Newton-Coulomb
theory, if the potentials $V$ and $\phi$ satisfy the
relation \eqref{mprelationGH1}, or equivalently, the relation
\eqref{Vp_chi}, then the pressure is given by 
\begin{equation}
 p =\epsilon\,\frac{\sqrt{G_{\rm d}}} {\beta}
\left(1-\beta^2\right) \, \int{\!\rho_{\rm
m}(\phi)\,d\phi} +p_0\, =  \epsilon\,\frac{\chi-\lambda}{1+\lambda} 
\sqrt{\frac{1+\lambda}{1+\chi}\,G_{\rm
d} }\, \int{\!\rho_{\rm m}(\phi)\,d\phi} + p_0 \, ,\label{pp_chi}
\end{equation}
$p_0$ being an integration constant, and in the case $\beta^2 = 1$, or
equivalently, in the case $\lambda=\chi$, the pressure is zero.
\end{cor}

\begin{proof}
 Consider the equilibrium equation in the form of Eq.~\eqref{eulerfunct},
and use the fact that from Theorem \ref{theoGH1} one has  $V' =
-\epsilon\,\beta\,\sqrt{G_{\rm d}\,}$ and 
$\rho_{\rm e}=\epsilon\,\beta\,\sqrt{G_{\rm d}}\, \rho_{\rm m}$, to find
$ p^\prime =    \left( \beta^2 -1 \right)\rho_{\rm e}
=  -\epsilon\,\frac{\sqrt{G_{\rm d}}} {\beta}
\left(\beta^2-1\right)\rho_{\rm m}$, which integrates to \eqref{pp_chi}.
\end{proof}

\noindent
In the case $\beta^2 = 1$ the pressure equals a constant, which in the
Newtonian theory is equivalent to zero pressure, in accordance to Theorem
\ref{theoNewt1}. This result corresponds to the Newtonian limit of the
relativistic relation between pressure and the metric potential for 
a charged fluid satisfying the Majumdar-Papapetrou relation for the
potentials, cf. Corollary \ref{theo-daray2-cor1}.

All of the fluid quantities can now be given in terms of only one variable,
the mass density, for instance, and with only one free parameter. Once the
mass density $\rho_{\rm m}$ is specified, the gravitational potential is
determined by the Poisson equation, $\nabla^2 \, V = S_{\rm
d-2}\,G\,\rho_{\rm m}$ and all of the other
quantities follow from $V$ and $\rho_{\rm m}$.

\subsection{The mass and charge, and the mass-charge relation of the global 
solution}

For completeness and for comparison with the relativistic theory let us
define mass and charge in the Newton-Coulomb theory. These quantities are
obtained by integration of the respective densities over the whole volume
of the source $\cal V_S$
\beqa
& & m = S_{\rm d-2}\int_{\cal V_S}  \rho_{\rm m}\, d{\cal V}\, ,\\
& & q = S_{\rm d-2}\int_{\cal V_S}  \rho_{\rm e}\, d{\cal V}\, .
\eeqa
For zero pressure, Eq.~\eqref{mpconditionp0} implies that the total
mass and total charge of the source are proportional to each
other. For Weyl type fluids with nonzero pressure, $m$ and $q$ are
proportional only in the case where $dp/d\phi$ is constant. In such a
case, one has the mass-charge relation
\beq
\beta\,q = \epsilon\sqrt{\,G_{\rm d}\,}\, m\,,
\label{masstocharge}
\eeq
which follows from Eq.~\eqref{chargerel1}.

\section{General relativistic charged fluid with pressure 
in ${\rm d}$ spacetime dimensions}
\label{sect-relatfluidmodel}

\subsection{Basic equations}
\label{sect-basic_relat}

Einstein-Maxwell equations in ${\rm d}$ spacetime dimensions are written as
\begin{eqnarray}
& &G_\munu= \frac{\rm d-2}{\rm d-3}\,S_{\rm d-2}G_{\rm d}\,
\left( T_\munu+ E_\munu\right)\, ,
\label{einst}\\
& & \nabla_\nu F^\munu = S_{\rm d-2}\,J^\mu\,, \label{maxeqs}
\end{eqnarray}
where Greek indices $\mu, \nu$, etc., run from $0$ to ${\rm d}-1$.
$G_\munu=R_\munu-\frac{1}{2}g_\munu R$ is the Einstein tensor, with
$R_\munu$ being the Ricci tensor,  $g_\munu$ the metric, and $R$ the Ricci
scalar. $S_{\rm d-2}$ and $G_{\rm d}$ have the same definitions as in the
Newton-Coulomb theory (See Sec. \ref{sect-basic_newton}, see also  
\cite{lemoszanchin2008}). We have put the speed of light equal to unity
throughout.  Note the singular behavior of the lower dimensional cases, 
${\rm d}=2$ and ${\rm d}=3$, which shall not be treated here. $E_\munu$ is
the electromagnetic energy-momentum tensor, which can be written as
\begin{equation}
S_{\rm d-2}\, E_\munu= {F_\mu}^\rho F_\nu{_\rho} -\frac{1}{4}g_\munu
F_{\rho\sigma} F^{\rho\sigma}\, ,\label{maxemt}
\end{equation}
where the Maxwell tensor is
\begin{equation}
F_\munu = \nabla_\mu A_\nu -\nabla_\nu A_\mu\, ,\label{ddemfield}
\end{equation}
$\nabla_\mu$ being the covariant derivative, and $A_\mu$ the
electromagnetic gauge field. In addition,
\begin{equation}
J_\mu = \rho_{\rm e}\, U_\mu\, ,\label{current}
\end{equation}
is the current density, $\rho_{\rm e}$ is the d-dimensional electric
charge density, and $U_\mu$ is the fluid velocity. $T_\munu$ is
the material energy-momentum tensor given by
\begin{equation}
T_\munu = \left(\rho_{\rm m}+p\right)U_\mu U_\nu
+p  g_\munu \, ,\label{fluidemt}
\end{equation}
where $\rho_{\rm m}$ is the fluid matter energy density in the
${\rm d}$-dimensional spacetime, and $p$ is the fluid pressure.

We assume the spacetime is static and that the metric can be written
in the form
\begin{equation}
ds^2 = - W^2 dt^2 + h_{ij}\,dx^idx^j\, ,
\quad\quad i,j=1,....,{\rm d}-1\,, \label{metricdd}
\end{equation}
a direct extension of the Majumdar-Papapetrou metric to extra dimensions.
The gauge field $A_\mu$ and four-velocity $U_\mu$ are then given by
\beqa
& &A_\mu = -\phi\,\delta_\mu^0\, ,\label{gauge1}\\
& &U_\mu =  -W\, \delta_\mu ^0\, .\label{veloc1}
\eeqa
The spatial metric tensor $ h_{ij}$, the metric potential $W$ and the
electrostatic potential $\phi$ are functions of the spatial coordinates
$x^i$ alone. Initially, we are interested in the equations determining the
metric potential $W$ and the electric potential $\phi$. These are obtained
respectively from the $tt$ component of Einstein equations (\ref{einst})
and from the $t$ component of Maxwell equations (\ref{maxeqs}). Such
equations give
\beqa
&& \nabla^2W = \,\frac{G_{\rm d}}{W}\left(\nabla_i\phi\right)^2
+S_{\rm d-2}\,G_{\rm d}
\,W\left(\rho_{\rm m}+ \frac{{\rm d}-1}{{\rm d}-3}\,p\right) \, ,
\label{ttein}\\
&& \nabla^2 \phi - \frac{1}{W}\nabla_i W\, \nabla^i\phi= -
S_{\rm d-2}\, W\,\rho_{\rm e}  \, , \label{max2}
\eeqa
where $\nabla_i$ denotes the covariant derivative with respect to the
coordinate $x^i$, with connection coefficients given in terms of the
metric $h_{ij}$.  

Eqs.~(\ref{ttein}) and (\ref{max2}) determine the potentials $W$ and $\phi$
in terms of a set of unknown quantities. Namely, the $({\rm d}-1)({\rm
d}-2)/2$ spatial metric coefficients $h_{ij}$ and the fluid variables,
energy density $\rho_{\rm m}$, electric charge density $\rho_{\rm e}$, and
pressure $p$. There are exactly $({\rm d}-1)({\rm d}-2)/2$ additional
equations that come from the Einstein equations, which in principle
determine the $h_{ij}$ metric components in terms of $\rho_{\rm m}$, $p$
and $\rho_{\rm e}$. Hence, to complete the system of equations it is
necessary to provide the energy and charge density functions, $\rho_{\rm
m}$ and $\rho_{\rm e}$, and also to specify the pressure $p$ or an equation
of state for the fluid. In the present analysis, we will not need the
explicit form of the space metric $h_{ij}$ and so the corresponding
Einstein equations will not be written here. Additional equations that can
be used are the conservation equations, $\nabla_\nu T^{\munu}=0$, which are
sometimes useful in replacing a subset of Einstein's equations. In the
present case the conservation equations yield
\beq
\left(\rho_{\rm m}+p\right)\frac{\nabla_i\, W}{W}
       +\rho_{\rm e}\,\frac{\nabla_i\,\phi}{W} +\nabla_i\,p
=0\label{conserveq}.
\eeq
This is the relativistic analogue to the Euler equation, and carries the
information of how the pressure gradients balance the equilibrium of the
system. It also shows that $p$, $W$ and $\phi$ are functionally related,
e.g., $p=p (W,\phi)$, or $W=W(p,\phi)$.

\subsection{Zero pressure: Weyl type systems in 
the Einstein-Maxwell theory and the De-Raychaudhuri
theorem in higher dimensions}
\label{sect-deraydd}

A special interesting case of charged matter is the dust fluid,
for which $p=0$ and Eq.~\eqref{conserveq} reduces to
\beq
\nabla_i\, W +\frac{\rho_{\rm e}}{\rho_{\rm m}} \,\nabla_i\,\phi\,
=0\, .\label{conserveq1}
\eeq
Given Eq.~(\ref{conserveq1}), we can render into ${\rm d}$ dimensions a
theorem stated in four dimensions, in part by De and Raychaudhuri
\cite{deraychaudhuri}, and fully by Guilfoyle \cite{guilfoyle} where
pressure is included. The theorem is the relativistic
version of our Newtonian theorem \ref{bonnor0}, which in turn
generalizes a theorem by Bonnor \cite{bonnor80}.

\begin{thm} {(De-Raychaudhuri 1968, Guilfoyle 1999)}
\label{theo-guilf261-relp0}

\item
For any charged static dust distributions in the Einstein-Maxwell theory,
the $({\rm d}-2)-$dimensional hypersurfaces of constant $W$ coincide
with the $({\rm d}-2)-$dimensional hypersurfaces of constant $\phi$,
and $W$ is a function of $\phi$ alone, $W = W(\phi)$.
\end{thm}

\begin{proof}
First observe that even though $W$ and $\phi$ are scalar functions in
a d-dimensional spacetime, since they do not depend upon time, each
one of the conditions of constant $W$ and $\phi$ indeed defines a
$({\rm d}-2)$-dimensional hypersurface. Moreover, from the
conservation equation \eqref{conserveq1} one gets $\rho_{\rm m} \,dW +
\rho_{\rm e}\,d\phi =0$. This equation implies that if any one of $dW$
or $d\phi$, are null, then both of them are. It also results in
$dW/d\phi = -\rho_{\rm e }/\rho_{\rm m}$ and the theorem follows.
\end{proof}
\noindent The result just stated ensures that in presence of static
dust fluid distributions the Weyl ansatz \eqref{weylhyp} is not
necessary, it is a consequence of the equilibrium equations.

It is now convenient to rewrite the preceding equations taking the
condition $W=W(\phi)$ into account. With this, Eqs.~(\ref{ttein}) and
(\ref{max2}) read, respectively,
\beqa
&&W^\prime \nabla^2\phi+\left(W''- \frac{G_{\rm d}}{W}\right)
 \left(\nabla_i\phi\right)^2  =  S_{\rm d-2}\,G_{\rm d}\,{W}
 \rho_{\rm m}, \label{ttein2a} \\
&&\nabla^2\phi -\frac{W'}{W}\left(\nabla_i \phi\right)^2 =
 -S_{\rm d-2}\,W \rho_{\rm e} \, ,\hfill\label{max3a}
\eeqa
where we have defined $W' = \displaystyle\frac{dW}{d\phi}$ and $W'' =
\displaystyle\frac{d^2 W}{d\phi^2}$. Substituting the functional relation
$W=W(\phi)$ into the conservation equations (\ref{conserveq1}) it follows
\beq
 W^\prime =-\frac{\rho_{\rm e}}{\rho_{\rm m}} \, .\label{conserveq1a}
\eeq
The basic system of equations to be solved is now composed of
Eqs.~\eqref{ttein2a}, (\ref{max3a}), and (\ref{conserveq1a}).  Such a
system can be thought of as determining the fluid variables $\rho_{\rm
m}$, and $\rho_{\rm e}$ once the potentials and metric functions are
known. On the other hand, $\rho_{\rm m}$ and $\rho_{\rm e}$ may be
eliminated from Eqs.~\eqref{ttein2a}, \eqref{max3a} and
\eqref{conserveq1a} to obtain
\begin{equation}
{\bar Z\,}\,\nabla_i\!\left({\bar Z\,}\nabla^i\phi\right)=0\,,
 \hfill\label{deraytheoeq1}
\end{equation}
where
\begin{equation}
\bar Z= \sqrt{{W^\prime}^2- G_{\rm d}\,}\, . \label{Zrelatp0}
\end{equation}
Eq.~\eqref{deraytheoeq1} has the same form of
Eq.~\eqref{bonnortheop0eq} (see \cite{deraychaudhuri} and also
\cite{bonnor80}). Hence the conditions of Theorem \ref{theoNewt1}
hold here, and we can render into d-dimensions a theorem initiated 
in part by Papapetrou \cite{papapetrou}, 
and stated in part by Das \cite{das62} and in
part by De and Raychaudhuri \cite{deraychaudhuri}.

\begin{thm}  {(Das 1962, De-Raychaudhuri 1968)}
\label{theo-deray1}
\item (i)
In the Einstein-Maxwell theory, 
if the surfaces of any charged dust distribution are closed equipotential
hypersurfaces and inside these hypersurfaces there are no holes or
alien matter, then the function $W(\phi)$ must satisfy the relation
\beq
W = - \epsilon\sqrt{G_{\rm d}}\, \phi\, +b  , \label{mprelationp0-rel}
\eeq
with $b$ being an integration constant, $\epsilon=\pm1$ as before,
and in such a case it follows
\beqa
& &\rho_{\rm e} = \epsilon\sqrt{G_{\rm d}\,}\, \rho_{\rm m}.
\label{mpconditionp0-rel}
\eeqa
\item (ii) 
In the Einstein-Maxwell theory, 
if in a spacetime region the ratio $\rho_{\rm e}/\rho_{\rm
m}$ equals a constant $K$, and there are no singularities, holes or alien
matter in that region, then the relation~\eqref{mprelationp0-rel} for the
potentials follows, and also $K= \epsilon\sqrt{{G_{\rm d}}}$.
\end{thm}

\begin{proof}
The proof of {\it (i)} is obtained by defining a new variable $\psi$,
such that $\nabla^i\psi = \bar Z\nabla^i\phi$, with $\bar Z$ given 
by Eq.~\eqref{Zrelatp0}. Following the same steps as in the case of the
Newtonian Theorem \ref{theoNewt1} {\it (i)} it can be shown that $\bar Z$
must be zero. Then Eqs.~\eqref{mprelationp0-rel} and
\eqref{mpconditionp0-rel} are obtained immediately.
The proof of {\it (ii)}  follows once we assume
that the ratio $\rho_{\rm e}/\rho_{\rm m}$ is
constant, and use Eqs.~\eqref{conserveq1a} and \eqref{Zrelatp0}.
Then $\bar Z=0$, resulting the same relations among $W$ and $\phi$,
and among $\rho_{\rm m}$ and $\rho_{\rm e}$ as in
Eqs.~\eqref{mprelationp0-rel} and \eqref{mpconditionp0-rel} above.
\end{proof}

\noindent
Papapetrou found the relation \eqref{mpconditionp0-rel} by assuming a
priori the relation between the potentials to be a perfect square,
$W^2 = (-\epsilon\sqrt{G\,}\phi+1 )^2$.  The theorem stated by Das
\cite{das62} assumes a priori that the level surfaces are the same
$W=W(\phi)$, which is not necessary according to
\cite{deraychaudhuri}, see also our theorems \ref{bonnor0} and 
\ref{theoNewt1} for the Newton-Coulomb theory.

\subsection{Nonzero pressure: Weyl type systems in 
the Einstein-Maxwell theory and new theorems in higher dimensions}
\label{sect-nonzerop-relat}

\subsubsection{Relativistic static charged pressure fluid: General results}
\label{sect-nonzerop-relat1}

As in the Newton-Coulomb case, Eq.~\eqref{conserveq} can be written in
terms of total derivatives. For contracting it with $dx^i$ yields
$\left(\rho_{\rm m}+p\right) \,(\nabla_i\, W) dx^i + \rho_{\rm
e}\,(\nabla_i\,\phi) dx^i +
W\left(\nabla_i\, p\right) dx^i =0$, or $\left(\rho_{\rm m}+p\right) dW+
\rho_{\rm e}\, d\phi + W\,  dp =0$. This implies that $W$, $\phi$ and $p$
are functionally related, and we can state a
theorem rendering into ${\rm d}$ spacetime dimensions a result
by Guilfoyle in ${\rm d} =4 $  \cite{guilfoyle}:

\begin{thm}
\label{theo-guilf261-rel} {(Guilfoyle 1999)}

\noindent
For any static charged pressure fluid in the Einstein-Maxwell theory, if any
two of the $({\rm d}-2)-$dimensional hypersurfaces of constant $W$,
$\phi$, or $p$ coincide, then the third also coincides. \end{thm}
\begin{proof} As shown above (see Theorem \ref{theo-guilf261-relp0}) the
conditions of constant $W$, $\phi$, or $p$ defines a $({\rm
d}-2)$-dimensional hypersurface. Moreover, from the conservation 
equation~\eqref{conserveq} 
one gets $\left(\rho_{\rm m}+p\right) \,dW + \rho_{\rm
e}\,d\phi+ W\,dp=0$. This equation implies that if any two of $dW$, $d\phi$,
or $dp$ are simultaneously null, then all the three are. This proves the
theorem.
\end{proof}

A further interesting consequence of the conservation equation is that,
thinking of $p$ as a function of $W$ and $\phi$, and observing that 
Eq.~\eqref{conserveq} implies in
$ dp = -\left(\rho_{\rm m}+p\right){dW}/{W} - \rho_{\rm  e}\,
{d\phi}/{W},$
we find 
$\left({\partial p}/{\partial W}\right)_{\phi} = - \,{\left(\rho_{\rm m}
+p\right)}/{W},\;\; {\rm and} \;\;\left({\partial p}/{\partial \phi}
\right)_{W} = -\, {\rho_{\rm e}}/{W}.$
Comparing to the Newton-Coulomb case, these relations confirm the fact that
in relativistic theory the pressure itself acts against the pressure
gradient, as the energy density, and that there are extra couplings between
energy and charge densities to the gravitational (metric) field.

\subsubsection{Weyl and Majumdar-Papapetrou relations for charged pressure
fluids in higher dimensions}
\label{sect-nonzerop-relat2}

Doing in $\rm d$-dimensional spacetimes what Weyl did in four-dimensional
general relativity \cite{weyl} even in the presence of matter with
pressure, we make the assumption on Einstein-Maxwell charged fluids in
d-dimensional spacetimes of being Weyl type fluids, where the metric
potential $g_{tt}\equiv -W^2$ is a functional of the gauge potential
$\phi$, $W=W(\phi)$, that we call here the Weyl ansatz. With such an
ansatz, Eqs.~(\ref{ttein}) and (\ref{max2}) read, respectively,
\beqa
&& W'\nabla^2\phi + \left(W''-\frac{G_{\rm d}}{W}\right)
\left(\nabla_i\phi\right)^2 = S_{\rm d-2}\,G_{\rm d}\,{W}
 \left(\rho_{\rm m}+ \frac{{\rm d}-1}{{\rm d}-3}\,p\right),
 \label{ttein2b} \\
&&\nabla^2\phi  - \frac{W'}{W}\left(\nabla_i \phi\right)^2=
 -S_{\rm d-2}\,W \rho_{\rm e} \, .\hfill\label{max3b}
\eeqa

For completion, and for comparison to the Newton-Coulomb case, let us
write here the result of the $\rm d$-dimensional
Weyl ansatz in vacuum (see \cite{lemoszanchin5,lemoszanchin2008}). Taking
into account the conditions $\rho_{\rm m}=0$, $p=0$, and $\rho_{\rm e}=0$,
Eqs.~\eqref{ttein} and \eqref{max2}  give $W\,W''+W '^2-G_{\rm d} = 0$,
i.e.,
\beq
W^2(\phi) =\left(-\epsilon\,\sqrt{G_{\rm d}\,}\phi +b\right)^2 +c
\label{weylrelationvacuum}\, ,
\eeq
where $b$ and $c$ are constant parameters. This is the $\rm d$-dimensional
version of the original Weyl relation, which is the analog of 
Eq.~\eqref{weylrelationNCvacuum} in the Newton-Coulomb theory.
For $c=0$ one gets, 
$
W= -\epsilon\,\sqrt{G_{\rm d}\,}\phi+ b\,
\label{mplrelationvacuum}
$
which is the Majumdar-Papapetrou relation, see 
Eq.~(\ref{mprelationp0-rel}).
We shall now work out the basic equations for relativistic matter with
nonzero pressure using the Weyl ansatz $W= W(\phi)$. From
Eq.~\eqref{conserveq} one has $p=(W,\phi)$ and, by assuming
further that $W$ is a function of $\phi$ alone, it follows from Theorem
\ref{theo-guilf261-rel} that the pressure
$p$ is also a function of the electric potential only, $p = p(\phi)$. 
Hence, Eq.~\eqref{conserveq} may be written as
\beq
(\rho_{\rm m}+p)\frac{W'} {W} + \frac{\rho_{\rm e}}{W}+ p^\prime =0
\, .\hfill \label{conserveq2b}
\eeq
We may now state more formally these results. In $ {\rm d} =4$ this
was shown by Guilfoyle \cite{guilfoyle}.

\begin{thm}  {(Guilfoyle 1999)}
\label{theo-guilf262-rel}

\noindent (i) If the Einstein-Maxwell charged pressure fluid is of Weyl
type and is in equilibrium, then the equipotentials are also hypersurfaces
of constant pressure, and vice-versa.
\item (ii) If the Einstein-Maxwell charged pressure fluid is of Weyl type
and is in equilibrium, then either
the pressure gradient vanishes at the surface of the fluid or the surface
is an equipotential.
\end{thm}

\begin{proof}
A relativistic Weyl type fluid satisfies the relativistic Weyl ansatz $W=
W(\phi)$ and Eq.~\eqref{conserveq2b} holds. From that equation we have
$(\rho_{\rm m}+p)d{W} +\rho_{\rm e}d\phi+ W dp =0$. Then since the
hypersurfaces of constant $W $ and $\phi$ coincide, Theorem
\ref{theo-guilf261-rel}
implies the third, the one of constant $p$ also coincides. Conversely,
a surface of constant $p$ implies $d\phi=0$, and thus it is also a surface
of constant $\phi$ and of constant $W$. This proves {\it (i)}.
To prove {\it (ii)} we first note that junction conditions establish that
the pressure is zero at the surface of the fluid, which means it is a
surface of constant pressure, $dp= (\nabla_i\, p)dx^i=0$. Then, unless the
pressure gradient vanishes at the surface, $\nabla_i p= 0$ for all $i$, by
{\it (i)} the surface of the fluid is an equipotential surface.
\end{proof}

We now go back to Eqs.~\eqref{ttein2b}-\eqref{conserveq2b}, 
and use them to eliminate $\rho_{\rm m}$ and
$\rho_{\rm e}$ in terms of the other quantities. The resulting equation,
that can be though of as an equation for $p$, may be cast into the form
\begin{equation}
{\bar Z\,}\nabla_i\!\left({\bar Z\,}\nabla^i \phi\right)\!
=-S_{\rm d-2}G_{\rm d}\,W^{2({\rm d}-2)/({\rm d}-3)}
\left(W^{-2/({\rm d}-3)}\,p\right)^\prime\, ,
\hfill\label{eqpprime-rel}
\end{equation}
where $\bar Z$ is given by~\eqref{Zrelatp0}. Usually, a relativistic
charged fluid problem is completely set out once we furnish the energy and
charge densities, and an equation of state for the fluid. However, as
it can be seen from Eq.~\eqref{eqpprime-rel}, for a Weyl type system,
the problem is in a position to be solved once the pressure gradient
$p'$ and the metric potential $W$ are given in terms of $\phi$. In
such a case, Eq.~\eqref{eqpprime-rel} can in principle be integrated
for $\phi$, from what all of the other fluid variables would
follow. We shall analyze some particular cases of this system next.

\subsubsection{New theorem in four and higher dimensions and the
Weyl-Guilfoyle relation: following Gautreau-Hoffman}
\label{sec-gautreau-rel}

Here we follow the approach by Gautreau and Hoffman \cite{gautreau} in
order to find the general properties of a charged pressure fluid
satisfying a Weyl-Guilfoyle relation, rather than the Weyl relation
alone as assumed in \cite{gautreau}. The results found in this section
generalize previous results in two ways. Fist, we render the theorem
given in \cite{gautreau} to higher dimensions, and, second, we find,
for ${\rm d}=4$ and ${\rm d}> 4$, 
the conditions that source matter distributions
must obey in order to satisfy the most general quadratic form for the
potentials $W^2 = a\,\left(-\epsilon\sqrt{G_{\rm d}}\phi+b\right)^2
+c$,
with arbitrary constants $a$, $b$, and $c$.

By defining the electromagnetic energy density as
\beq
\rho_{\rm em } = \frac{1}{S_{\rm
d-2}}\frac{\left(\nabla_i\phi\right)^2}{W^2}\,, \label{emdensity}
\eeq
one may cast Eqs.~\eqref{ttein} and \eqref{max2} into the form
\beqa
&& \nabla^2 W = S_{\rm d-2}\,G_{\rm d}
\,W\left(\rho_{\rm m}+ \frac{{\rm d}-1}{{\rm d}-3}\,p +
\rho_{\rm em}\right) \, , \label{tteinGH1}\\
&& \nabla_i\left(\frac{1}{W}\,\nabla^i\phi\right) = -
S_{\rm d-2}\, \rho_{\rm e}  \, . \label{maxGH1}
\eeqa
Also, using the identities $\nabla_i\left(\frac{1}{W}\,\nabla^i W^2\right) 
= 2\nabla^2 W$ and $\nabla_i\left(\frac{1}{W}\,\nabla^i\phi^2\right)=
2\frac{\phi}{W}\nabla^2\phi- 2\frac{\phi}{W^2}\nabla_i W\nabla^i\phi+
2{S_{\rm d-2}}{W} \rho_{\rm em}$, we see that the Eqs.~\eqref{tteinGH1}
and \eqref{maxGH1} may be rearranged as
\beqa
&&\nabla_i\left(\frac{1}{W}\,\nabla^i W^2\right)= 2S_{\rm d-2}\,G_{\rm d}
\,W\left(\rho_{\rm m}+ \frac{{\rm d}-1}{{\rm d}-3}\,p +
\rho_{\rm em }\right) \, , \label{tteinGH2}\\
&&    \nabla_i\left(\frac{1}{W}\,\nabla^i\phi^2\right) = - 
2S_{\rm d-2}\left(
\rho_{\rm e}\phi - \rho_{\rm em }{W}\right)  \, . \label{maxGH2}
\eeqa
Then, multiply~\eqref{maxGH1} by $-2ab\epsilon\sqrt{G_{\rm d}}$ and add
to \eqref{maxGH2} multiplied by $a{G_{\rm d}}$, with constant $a$
and $b$, and subtract the result from Eq.~\eqref{tteinGH2} to find 
\beqa
\nabla_i\left(\frac{1}{W}\nabla^i\left[W^2 -a\left(-\epsilon\sqrt{G_{\rm d}\,}
\phi +b\right)^2 -c\right]\right) = & 2 S_{\rm d-2}G_{\rm d} & \left[
\left(\rho_{\rm m}+ \frac{{\rm d}-1}{{\rm d}-3}\,p
+\rho_{\rm em} \right)W + \right. \nonumber\\
&&+ \left.a\left(\phi\rho_{\rm e} -W\rho_{\rm em}\right)
-\epsilon\,\frac{ab}{\sqrt{G_{\rm d}\,}}\rho_{\rm e}  \right] .
\label{finaleqGH}
\eeqa
On basis of this equation some interesting conclusions can be drawn.

\begin{thm} {(New)}
\label{theoGH1-rel}

\noindent
\item (i) In any Einstein-Maxwell charged pressure fluid, if
the potentials are such that $\,W^2 - a \left(-\epsilon\sqrt{G_{\rm
d}}\phi+b\right)^2-c$
vanishes everywhere, i.e., if $W^2= a \left(-\epsilon\sqrt{G_{\rm
d}}\phi+b\right)^2+c$, then the charged pressure fluid quantities satisfy
the constraint
\beq
 a\,b\, \rho_{\rm e}  =\epsilon\sqrt{G_{\rm d}}\left[\left(\rho_{\rm m}+
\frac{{\rm d}-1}{{\rm d}-3}\,p \right)W +\phi\rho_{\rm e}
+\left(a-1\right)\left(\phi\rho_{\rm e} -  W \rho_{\rm em}\right)\right] \,,
\label{mpconditionGH1}
\eeq
or
\beq
 a\, \rho_{\rm e}\left(-\epsilon\sqrt{G_{\rm d}}\phi +b\right)  =
\epsilon\sqrt{G_{\rm d}}\left(\rho_{\rm m}+ \frac{{\rm d}-1}{{\rm d}-3}\,p
+ \epsilon\left(1-a\right)\rho_{\rm em}\, W \right) \,,
\label{mpconditionGH2}
\eeq
which can be considered the equation of state satisfied by the charged
fluid.

\noindent
(ii) Conversely, in any Einstein-Maxwell charged pressure fluid, if
the fluid quantities are such that Eq.~\eqref{mpconditionGH1} holds and
there is a closed surface, with no singularities, holes or alien matter
inside it, where $W^2 - a \left(-\epsilon\sqrt{G_{\rm d}}\phi+b\right)^2
-c$ vanishes, then
\beq
W^2 = a \left(-\epsilon\sqrt{G_{\rm d}}\phi+b\right)^2+c\,
\label{weylrelationGH1}
\eeq
everywhere inside the surface.
\end{thm}

\begin{proof}
The proof of this theorem is obtained in the exact same way as for the
Newton-Coulomb case of theorem \ref{theoGH1}.
\end{proof}

\noindent
Theorem \ref{theoGH1-rel} establishes that the generalized quadratic
Weyl-Guilfoyle relation, i.e., Eq.~\eqref{weylrelationGH1} with $a\neq
1$, leads to a constraint among the fluid quantities which includes
the electromagnetic energy density $\rho_{\rm em}$ and also the
binding energy density $\phi\rho_{\rm e}$.  It generalizes the
analysis by Gautreau and Hoffman \cite{gautreau} in two ways. First it
holds for arbitrary values of constant $a$, while in
 \cite{gautreau} $a=1$, and second it holds in an arbitrary number of
spacetimes dimensions $d\geq 4$, rendering into higher dimensions the
results of Gautreau and Hoffman \cite{gautreau}.  When $a$ equals
unity, the ${\rm d}$-dimensional Weyl relation is recovered and
Eq.~\eqref{mpconditionGH1} is reduced to the form found in
\cite{lemoszanchin5}. Also, in ${\rm d}=4$ it results in
what was found by Gautreau and Hoffman \cite{gautreau}. Comparing our
theorem to the theorem by Gautreau and Hoffman we see that in their
analysis only the binding energy was taken into account, and hence the
constant $a$ was forced to be equal to unity.
Furthermore, all the solutions found by Guilfoyle in \cite{guilfoyle} 
obey the equation of state, a constraint, provided by 
Eq.~\eqref{mpconditionGH1}.

The functional form \eqref{weylrelationGH1} is more general than the
original quadratic Weyl type form. As pointed out by Guilfoyle
\cite{guilfoyle}, a reparameterization of the metric potential as $W
\longrightarrow \sqrt{a\,} W $, with $a> 0$ enables us to write
Eq.~\eqref{weylrelationGH1} as $W^2= \left(-\epsilon\sqrt{G_{\rm
d}}\phi+b\right)^2 +c/a$, which is indeed the original Weyl quadratic
form. However, one needs to observe that this reparameterization,
which corresponds to rescaling the time coordinate as
$t\longrightarrow t/\sqrt{ a}$, leads to a new system where, the mass
density, the pressure and the electric charge density of the fluid are
rescaled as $\rho_{\rm e} \longrightarrow \rho_{\rm e}/{ a}$, $p
\longrightarrow p/a$, and $\rho_{\rm e} \longrightarrow \rho_{\rm
e}/{\sqrt a}$, respectively, which changes the balancing relation
among the mass and charge densities and the pressure gradient. We
explore this fact in the next section.

\subsubsection{The same new theorem as last subsection: following
De-Raychaudhuri}
\label{newthm-rel}

We shall analyze here the Weyl type charged pressure fluid now
following the strategy of De and Raychaudhuri \cite{deraychaudhuri},
in which the conservation equation is used to try to find a
differential equation for the potentials, by eliminating all the fluid
quantities from the system. This strategy has already been adopted in
the Newton-Coulomb case of Sec.~\ref{sect-deraynewton}. The strategy
is different from the Gautreau-Hoffman one, but the results are
similar.  Nevertheless it is interesting to see where this strategy
leads to.

A comparison between Eqs.~\eqref{eqpprime} and \eqref{eqpprime-rel}
reveals that the quantity $W^{-2/({\rm d}-3)}\,p$ plays, in the
Einstein-Maxwell with matter theory, a similar role to the one played by
the fluid pressure $p$ alone in the Newton-Coulomb with matter theory.
Hence, in order to write the relativistic equations in a form that
resembles the Newton-Coulomb case, we shall define an effective matter
density $\bar\rho_ {\rm m}$ and an effective pressure $\bar p$ respectively
by
$\bar {\rho}_{\rm m} = \rho_{\rm m}+
\frac{{\rm d}-1}{{\rm d}-3}\,p\,,$ and $\bar p = W^{-2/({\rm
d}-3)}\,p. $ With these definitions, Eq.~\eqref{conserveq2b} assumes
the form
$
\bar {\rho}_{\rm m} W^\prime + {\rho}_{\rm e}+
 W^{\frac{{\rm d}-1}{{\rm d}-3}}\, {\bar {p}\,}^\prime =0 \, .
$
This result suggests that the product of the effective pressure
gradient $\bar p\,'$ and some power of the metric potential $W$, such
as $W^{\delta}\,p$, for some number $\delta$, plays the role of the
charge density analogously to the Newtonian case. Notice, however,
that in the relativistic case the electromagnetic energy density $\sim
\left(\nabla_i\phi\right)^2 $ is also source to the gravitational
field $W$ (cf. Eq.~\eqref{ttein}). Then, it is seen that the pressure
gradient ${\bar{p}}^\prime$, besides being connected to the charge and
mass densities, is also connected to the electromagnetic
energy density. In fact, after a careful analysis, one finds that by
defining two new quantities $\bar\lambda$ and $\bar\chi$ through the
relation
$
W^{\frac{{\rm d}-1}{{\rm d}-3}}\,{\bar p}^\prime =\frac{\bar\chi\,}{W^2}
\rho_{\rm e}+ {\bar\lambda} W^\prime\left(\bar\rho_{\rm m} +
\rho_{\rm em}\right),
$
with $\rho_{\rm em}$ defined in Eq.~\eqref{emdensity},
it is possible to put Eq.~\eqref{eqpprime-rel} into a similar form to 
Eq.~\eqref{eqforchi}. Going back to the original
quantities, ${\rho}_{\rm m}$, $p$, ${\rho}_{\rm e}$, and $\rho_{\rm em}$,
we find the equation
\beq
\left(p^\prime - \frac{2}{{\rm d}-3}\frac{W^\prime}{W}p\right)W=
 \frac{\bar\chi}{W^2 }\,\rho_{\rm e} +\bar\lambda {W^\prime}
\left(\rho_{\rm m}+ \frac{{\rm d} -1}{{\rm d}-3}\,p +
\rho_{\rm em}\right)\,, \label{pgradient-rel}
\eeq
which is similar in form to the corresponding equation, Eq.
\eqref{pgradient}, of the Newtonian theory. Here $\bar\lambda$
and $\bar\chi$ are arbitrary functions of $\phi$ alone. Then,
Eq.~\eqref{eqpprime-rel} assumes the form
\beq
\bar Z_{p}\nabla_i\left(\bar Z_{p} \nabla^i\phi\right)  = \frac{G_{\rm
d}}{2}
\left({W^\prime}^2 \bar\lambda^\prime - \frac{\bar\chi^\prime}{W^2}\right)
\left(\nabla_i\phi\right)^2\, ,\label{eqforchi-rel}
\eeq
where we have defined
\beq
\bar Z_{p}= \sqrt{(1+\bar\lambda)W^\prime\,^2-\,
\left({1+\frac{\bar\chi}{W^2}}\right)\,G_{ \rm d}} \, .\label{Zrelp_chi}
\eeq
As mentioned above, the study done in the present section was inspired in
the work by De and Raychaudhuri \cite{deraychaudhuri} on the relativistic
charged dust fluid in four-dimensional spacetimes. Besides performing the
analysis in arbitrary dimensions, we have also generalized the analysis by
including nonzero pressure and found that the pressure gradient plays the
role of electric charge density. Then Eq.~\eqref{eqforchi-rel} is the
fundamental equation to be solved for $\phi$, 
noting that it is still needed to
furnish the functions $W(\phi)$, $\bar\lambda(\phi)$ and $\bar\chi(\phi)$.
There is, however, a particularly interesting case that deserves further
analysis. In fact, as in the Newton-Coulomb case, for constant
$\bar\lambda$ and $\bar\chi$, results that are equivalent to those
stated in the theorem \ref{theoGH1-rel} can be found.

\begin{thm}  (Same as theorem \ref{theoGH1-rel}, following another path)
\label{theo-deray2}

\noindent
(i) If the surfaces of any Einstein-Maxwell charged pressure fluid
distribution are closed equipotential hypersurfaces and inside these
hypersurfaces there are no singularities, holes or alien matter, and the
fluid is of Weyl type, whose pressure satisfies the condition given in
Eq.~\eqref{pgradient-rel} with constant $\bar\lambda$ and $\bar\chi$, then
the function $W(\phi)$ must satisfy the relation
\beq
W^2  = \frac{1}{1+\bar\lambda }\left(-\epsilon\,\sqrt{G_{\rm d}}\,\phi
+b\right)^2 -\bar\chi\, ,\hfill \label{weylrelationp_chi-rel}
\eeq
$b$ being an arbitrary integration constant, and it follows
\beq
b\rho_{\rm e}  = \epsilon\left(1+\bar\lambda\right)\sqrt{G_{\rm d}}\,
\left[\left(\rho_{\rm m}+\frac{\rm d-1}{\rm d-3}\, p\right)W
+\phi \,\rho_{\rm e} -\frac{\bar\lambda}{1+\bar\lambda}\,\left(\phi
\,\rho_{\rm e} - W \rho_{\rm em}\right)\right]
 \, , \label{mpconditionp_chi-rel}
\eeq
or equivalently
\beq
\rho_{\rm e}\left(-\epsilon\sqrt{G_{\rm d}}\phi+b\right)  =
\epsilon\,\left(1+\bar\lambda \right)\sqrt{G_{\rm d}}\,
\left(\rho_{\rm m}+\frac{\rm d-1}{\rm d-3}\, p
+\frac{\bar\lambda}{1+\bar\lambda }\, \rho_{\rm em}\right)W  \, .
\label{mpconditionp_chi-relb}
\eeq

\noindent
(ii) If condition \eqref{mpconditionp_chi-rel} holds everywhere inside the
Einstein-Maxwell charged pressure fluid, with constant $\bar\lambda$ and
$\bar\chi$, and there are no singularities, holes or alien matter in that
region, then the function $W$ is of the form given in
Eq.~\eqref{weylrelationp_chi-rel}.
\end{thm}

\begin{proof}
The proof of {\it (i)} may be given by defining a new variable $\psi$ by
$\nabla_i\psi = \bar Z_{p}\nabla_i\phi$, with $\bar Z_{p}\nabla_i\phi$
given by \eqref{Zrelp_chi}, and then following the same steps as done in
the case of Theorem \ref{theoNewt1}. In the end, it is found that 
$\bar Z_{p}\nabla_i\phi$ vanishes everywhere in the fluid which means
$\sqrt{1+\bar\lambda}\,W^\prime =
 -\epsilon\sqrt{(1+\bar\chi/W^2)G_{\rm d}}.$
By integrating this equation one gets the relation
\eqref{weylrelationp_chi-rel}, and then using this result and
Eqs.~\eqref{conserveq2b} and \eqref{pgradient-rel} one gets the constraint
\eqref{mpconditionp_chi-rel}. On the other hand, part {\it (ii)} follows
straightforwardly from Eq.~\eqref{mpconditionp_chi-rel},
Eqs.~\eqref{pgradient-rel} and \eqref{eqforchi-rel}.
\end{proof}

\noindent
Even though we have followed different strategies in each case, the results
stated in theorem \ref{theo-deray2} are equivalent to what is stated in
theorem \ref{theoGH1-rel}. Namely, the obtained relations among fluid
quantities and potentials given in Eqs.~\eqref{weylrelationp_chi-rel} and
\eqref{mpconditionp_chi-rel} (or \eqref{mpconditionp_chi-relb}) are
equivalent to the relation given by Eqs.~\eqref{weylrelationGH1} and
\eqref{mpconditionGH1}, respectively. This can be shown explicitly through
the appropriate identifications of the arbitrary constants, namely, $ a =
1/(1+\bar\lambda)$ and $c= -\bar\chi$.
Let us now comment on three interesting particular cases: (i) $\bar\chi
=0$, $\bar\lambda \neq 0$, (ii) $\bar\chi \neq 0$, $\bar\lambda=0$ and
(iii) $\bar\chi \neq 0$ and $\bar\lambda= 0$. In the case (i) $\bar\chi
=0$, for arbitrary $\bar\lambda$, the metric potential is a perfect square
function of the electric potential, $W =  \left(-\epsilon\,\sqrt{G_{\rm
d}}\, \phi +b\right)/\sqrt{(1+\bar\lambda)} $. This is closely related to
the Newton-Coulomb case studied in Sec.~\ref{sect-deraynewton}, as it can
be seen from its Newtonian limit. Case (ii) corresponds to the Weyl
original quadratic form between the potentials, for which several studies
in ${\rm d}=4$ have been done. The particular case (iii) in which both of
the free parameter are zero, $\bar\lambda =0$ and $\bar\chi= 0$
reproduces the result found in  \cite{lemoszanchin5}, and in 
\cite{guilfoyle} in four-dimensional spacetimes. This is a special case
where the conditions for Theorem \ref{theo-deray1} hold even for nonzero
pressure.

 As seen above, Eq.~\eqref{eqpprime-rel} is the fundamental
equation and its validity is guaranteed also in the case $\bar\chi
=\bar\lambda=0$. In fact, from Eq.~\eqref{pgradient-rel} one gets
\begin{equation} W^{2/({\rm d}-3)} \left(W^{-2/({\rm d}-3)}p\right)^\prime=
 p^\prime -\frac{2}{{\rm d}-3}p\,\frac{W^\prime}{W} = 0 \, ,
\label{pWcondition} \end{equation} and then the right-hand side of Eq.
\eqref{eqpprime-rel} vanishes identically so that the conditions of Theorem
\ref{theo-deray1} hold. In other words, if $\bar\lambda=0$ and $\bar\chi=0$
the Majumdar-Papapetrou relation holds. So one may state the following
corollary, which was shown as a theorem for four-dimensional spacetime by
Guilfoyle \cite{guilfoyle}, see also Lemos and Zanchin
\cite{lemoszanchin5} for the d-dimensional generalization. We repeat it
here for completeness and because it follows as corollary of the previous
theorems, rather than being a theorem itself.

\begin{cor}  {(Guilfoyle 1999, Lemos-Zanchin 2005)}
\label{theo-daray2-cor1}

\noindent
In a region of a static spacetime, filled by a charged pressure fluid of
Weyl type, the relation
\beq
p= k W^{2/({\rm d}-3)}\, , \label{pguilfoyle}
\eeq
 holds if and only if
\beq
W= -\epsilon\sqrt{G_{\rm d}}\phi+ b,
\label{mprelation-rel1}
\eeq
and there is a closed equipotential surface in that region of the
spacetime with no singularities, holes, or alien matter inside it. In such
a
case cit follows the relation
\beq
\rho_{\rm e} =\epsilon\sqrt{G_{\rm d}\,}\left(\rho_{\rm m} 
+ \frac{{\rm d}-1} {{\rm d}-3}p\right). \label{mpconditionguilfoyle}
\eeq
\end{cor}

\begin{proof}
The proof is given by observing that if the relation \eqref{pguilfoyle}
holds, the right-hand side of Eq.~\eqref{eqpprime-rel} vanishes, and
then from Theorem \ref{theo-deray1} the results~\eqref{mprelation-rel1} and
\eqref{mpconditionguilfoyle} follow. On the contrary, if
\eqref{mprelation-rel1} holds, Eq.~\eqref{pguilfoyle} follows directly from
Eq.~\eqref{eqpprime-rel}, there is no need of a closed surface here, and
the result given in Eq.~\eqref{mpconditionguilfoyle} follows from
Eq.~\eqref{mpconditionp_chi-relb} with $\bar\lambda =0$. \end{proof}

\noindent
The Newtonian limit of this relativistic solution is a pressureless
fluid, as will be shown below.

\subsubsection{Spherically symmetric spacetimes}
\label{sphericalsystems}

The presence of the constant $\bar\lambda \neq 0$, or, in the notation
of Sec. \ref{sec-gautreau-rel}, the presence of constant $a \neq 1$,
in the formula for the potentials is related to the work by Guilfoyle
\cite{guilfoyle} in four dimensional spacetimes. In fact, it can be
shown that the Weyl type solutions found in \cite{guilfoyle} satisfy
the conditions given by Eqs.~\eqref{mpconditionGH1} and
\eqref{weylrelationGH1}, or equivalently, by
Eqs.~\eqref{weylrelationp_chi-rel} and
\eqref{mpconditionp_chi-relb}. To show that, and also to verify that
for ${\rm d}=4$ our results are consistent with previous work,
let us analyze one of the spherical solutions of \cite{guilfoyle}. The
spacetime metric, in Schwarzschild coordinates, is taken in the form
$ds^2 = -W^2(r)\,dt^2 +dr^2/\sqrt{1 -r^2/R^2\,} + r^2\, d\Omega^2$,
with $r$ being the radial coordinate and $R$ being a constant. Take, for
instance, the solution found in \cite{guilfoyle} for $A=3/2$, which
corresponds to $\bar\lambda = 1/2$, or $a=2/3$, according to our
notation,
\beqa
 W^2(r) & = &\frac{3\,\bar\chi^3 \, F^2(r)}{3\,\bar\chi^2 F^2(r)-2} \, ,
    \label{W-sol}  \\
 8\pi   \rho_{\rm m}(r)& =& \frac{3}{R^2} - \frac{9\,\bar\chi^2\,k^2r^2}
{\left(3\,\bar\chi^2\, F^2(r)-2\right)^2}\, , \label{rhom-sol} \\
Q(r) &=&  \frac{-3\epsilon\, \bar\chi\, k\, r^3}{3\,\bar\chi^2\, F^2(r)-2}
       , \label{charge-sol}\\
 8\pi p(r) &=& -\frac{1}{R^2} + \frac{9\,\bar\chi^2\, k^2r^2}
{\left(3\,\bar\chi^2\, F^2(r)-2\right)^2}-\frac{4k\sqrt{1-\frac{r^2}{R^2}\,}}
      {F(r)\left(3\,\bar\chi^2\,F^2(r)-2\right)}  , \label{p-sol}
\eeqa
where $k$ is an integration constant, and $F(r)$ and $Q(r)$ are defined by
\beqa
F(r) &=& c_0 - kR^2\sqrt{1 - \frac{r^2}{R^2}}\, ,\\
Q(r) & =& 4\pi\int_0^r \rho_{\rm e}(r)
\frac{r^2 dr}{\sqrt{1-\frac{r^2}{R^2}}}= \frac{r^2}{W} \sqrt{1
-\frac{r^2}{R^2}} \left(\frac{d\phi(r)}{dr}\right)^2  ,
\label{charge-def}
\eeqa
with $c_0$ being another integration constant, and we have put 
$G_4\equiv G=1$.

From Eqs.~\eqref{charge-def} and \eqref{charge-sol} we get both the electric
charge density $\rho_{\rm e}$  and the electromagnetic energy density
$\rho_{\rm em}$,
\beqa
8\pi  \rho_{\rm e}(r) & =& \frac{-18\epsilon
\,\bar\chi k}{3\,\bar\chi^2\,F^2(r)-2}  \left(\sqrt{1- \frac{r^2}{R^2\,}} -
\,\frac{2\,\bar\chi^2k\, r^2 \, F(r)}
{\,3\,\bar\chi^2\, F^2(r)-2}\right)  , \label{rhoe-sol}\\
8\pi  \rho_{\rm em}(r)  & =&
\frac{18\,\bar\chi^2 k^2r^2}{\,\left(3\,\bar\chi^2\,F^2(r)-2\right)^2}
\, . \label{rhoem-sol}
\eeqa
In the case ${\rm d}=4$ and $\bar\lambda=1/2$,
Eq.~\eqref{mpconditionp_chi-relb} reads
\beq
\rho_{\rm e}  = \sqrt{\frac{3}{2}}
\left(\rho_{\rm m}+ 3 p +\frac{\rho_{\rm em}}{3} \right)
\frac{\epsilon\, W} {\sqrt{W^2+\bar\chi\,}}\, . \label{mpconditionp_chi-sol}
\eeq
Hence, in order to check if the solution satisfies this constraint, we use
Eqs.~\eqref{rhom-sol} \eqref{p-sol}, and \eqref{rhoem-sol} to obtain
\beq
8\pi \sqrt{\frac{3}{2}} \left(\rho_{\rm m}+ 3p +\frac{\rho_{\rm
em}}{3}\right) = -\frac{12 k\sqrt{3/2}
}{3\,\bar\chi^2\,F^2(r)-2}\left(\frac{1}{F(r)}\sqrt{1-\frac{r^2}{R^2}} -
\frac{2\,\bar\chi^2 k r^2} {\, 3\,\bar\chi^2\,F^2(r)-2}\right)\,,  
\eeq
where we have multiplied both sides of the last equation by $8\pi$.
Now, from Eq.~\eqref{W-sol} we find
$
\frac{W} {\sqrt{W^2+\bar\chi\,}} = \sqrt{\frac{3}{2}}\,\bar\chi
F(r)\,, 
$
and then the right hand side of Eq.~\eqref{mpconditionp_chi-sol}
corresponds to
\beq
 8\pi\sqrt{\frac{3}{2}} \left(\rho_{\rm m}+ 3p +\frac{\rho_{\rm
 em}}{3}\right) \frac{\epsilon\,W} {\sqrt{W^2+\bar\chi\,}}
=-\frac{18\epsilon\,\bar\chi k}{3\,\bar\chi^2\,F^2(r)-2}
\left(\sqrt{1-\frac{r^2}{ R^2}} -
\frac{2\,\bar\chi^2 k r^2 F(r)} {\, 3\,\bar\chi^2\,F^2(r)-2}\right)\,,
\eeq
agreeing with the solutions for $\rho_{\rm e}(r)$ as in
Eq.~\eqref{rhoe-sol}, and showing that this particular solution by
Guilfoyle \cite{guilfoyle} really matches the conditions established
in the present work.  With confidence we can state that all
Guilfoyle's solutions \cite{guilfoyle} obey the condition
(\ref{mpconditionGH1}) or the other equivalent ones.

\subsubsection{The Newtonian limit}
\label{secNewtonianlimit}

The Newtonian limit can be obtained from the relativistic quantities by
 considering Eq.~\eqref{eqpprime-rel} with the following approximations:
(i) Write $W = 1+F$ and consider
$F=F(\phi)$ small compared to unity, $F$ is to be compared to the Newtonian
potential $V$; (ii) Neglect $p$ with respect to $\rho_{\rm m}$; (iii) The
product $p^\prime W=p^\prime(1+F)$ is approximated by $p^\prime$ since
$p^\prime F$ is also a second order term; (iv) Neglect $pW^\prime$ when
compared to $p^\prime W$, because it gives $pW^\prime= pF^\prime$ which is
a second order term too; (v) The product  $\rho_{\rm e} W  = 
\rho_{\rm e}(1+F)$ is approximated to $\rho_{\rm e}$. It then follows that
$ W p^\prime  -\frac{1}{{\rm d}-3}\,p W^\prime \simeq p^\prime
-\frac{1}{{\rm d}-3}\, p F^\prime\, $, and from
Eq.~\eqref{Zrelatp0}, that $ \bar Z = {F^\prime}^2 - G_{\rm d}$.
Therefore,
 Eq.~\eqref{eqpprime-rel} reduces to
\beq
\sqrt{{F^\prime}^2-G_{\rm d} \,}\nabla_i\!\left(\sqrt{{F^\prime}^2
-G_{\rm d}\,}\nabla^i \phi\right)\!\! = S_{\rm d-2}\,G_{\rm d}\, 
 p^\prime \, ,\hfill
\label{basiceq-Newton}
\eeq
which, after the identifications $F = V$ and 
$\bar Z= Z=\sqrt{{F'}^2- G_{\rm d}}$
gives exactly Eq.~\eqref{bonnortheop0eq}, that is the basic equation
for the Newton-Coulomb theory (see also \cite{bonnor80}). From this
result one then finds the corresponding results for zero pressure, cf.
Eqs.~\eqref{mprelationp0} and \eqref{mpconditionp0}. The Newtonian
results for nonzero pressure in which $p^\prime = \chi \rho_{\rm
e}+\lambda V^\prime \rho_{\rm m} $, with constant $\chi$ and
$\lambda$, also follow immediately (cf. Eqs.~\eqref{Vp_chi},
\eqref{rhop_chi} and \eqref{pp_chi}).

Alternatively, one may obtain the Newtonian limit directly from the
relativistic solutions found above. For instance, considering the case of
Sect. \ref{newthm-rel} the Newtonian limit can be found by writing 
Eq.~\eqref{weylrelationGH1} in the form $W^2=  a\,G_{\rm d}\phi^2 
-2\epsilon\,a\,b\,\sqrt{G_{\rm d}}\, \phi + a\, b^2 +c$, so that to first
order approximation in $\phi$ one gets $W^2= a\, b^2 +c -
2\epsilon\,a\,b\,\sqrt{G_{\rm d}}\, \phi$. Now writing $W \simeq 1
+2V$ and redefining appropriately the constants $a$, $b$ and $c$ in terms of
$\lambda$, $\gamma$ and $\chi$, Eq.~\eqref{Vp_chi} follows. Also, the two
remaining Newtonian relations, Eqs.~\eqref{rhop_chi} and \eqref{pp_chi},
are immediately obtained as the Newtonian limit of the corresponding
relativistic equations. Namely, Eq.~\eqref{rhop_chi} follows from
Eq.~\eqref{mpconditionp_chi-rel}, with $W^2 \sim 1$ and neglecting the
pressure $p$ when compared to the matter density $\rho_{\rm m}$, and
Eq.~\eqref{pp_chi} is obtained after
integration of the Newtonian approximation of Eq.~\eqref{pgradient-rel},
which is $ p^\prime = \bar\chi \rho_{\rm e}+\bar\lambda V^\prime \rho_{\rm
m} $.  Note that the relation among the constants
$\beta=\sqrt{(1+\chi)/(1+\lambda)}$ and $\gamma$ used in the Newton-Coulomb
theory and the constants $a = 1/(1+\bar\lambda)$, $b$ and $c=-\bar\chi$
used in the Einstein-Maxwell theory is not unique, since it depends on how
the Newtonian limit is calculated. By taking, $W^2 = 1 + 2V$, we find
$\beta= a\, b$, and $\gamma= (ab^2+ c-1)/2$.

Additionally, the relativistic relation $p=k W^{2/({\rm d}-3)}$ found in
Sect. \ref{newthm-rel}, which to first order approximation is $p=k=$
constant, and so $p'=0$, corresponds to the zero pressure Newton-Coulomb
fluid studied in Sect. \ref{sect-bonnordd} (cf. Corollary \ref{cor-newt}).
In fact, in the Eulerian description of a Newton-Coulomb charge fluid the
conditions $p=0$ and $p=$ constant are equivalent.

\subsection{Asymptotically Tangherlini spacetimes: the mass and charge, 
and the mass-charge type relation of the global solution}
\label{sect-massdef}

Here we particularize the analysis to asymptotically Tangherlini
spacetimes, i.e., Reissner-Nordstr\"om $\rm d$-dimensional spacetimes,
filled by charged fluids of Weyl-Guilfoyle type. Such spacetimes have
an asymptotically electrovacuum region, which implies strong
constraints on the fluid distributions. Namely, the source fluid has
to be of finite extent with a well defined boundary, or the fluid
quantities must approach zero in a sufficiently fast form, in such a
way to guarantee the existence of an asymptotically vacuum region.  We
show below that in the case the fluid distribution has no definite
boundary, or whenever a Weyl-Guilfoyle type relation of the form $W^2
= a\, \left(-\epsilon\,\sqrt{G_{\rm d}}\,\phi+b\right)^2 +c$ is
imposed in the whole spacetime, the existence of an asymptotically
electrovacuum region requires that $a=1$, $c+ b^2=1$ throughout, and
the mass and electric charge are related by $b\, q
=\epsilon\,\sqrt{G_{\rm d}}\,m$.

The case of a fluid distribution with a boundary is more interesting,
since different Weyl-Guilfoyle type relations may be assumed in each
region of the spacetime. In fact, solutions of this kind were studied
by Guilfoyle himself \cite{guilfoyle}, who had found a more general
mass to charge relation. Below we extend the Guilfoyle analysis to
$\rm d$-dimensional spacetimes.  Following Gautreau and Hoffman
\cite{gautreau} we also give a more formal definition of mass and
charge for $\rm d-$dimensional asymptotically Reissner-Nordstr\"om
spacetimes of Weyl-Guilfoyle type. Finally, we show that in
asymptotically Reissner-Nordstr\"om spacetimes the mass definition
given here is identical to the Tolman mass \cite{tolman}.

\subsubsection{Spherically symmetric interiors
joined to an exterior Tangherlini spacetime: relation
between the mass, charge and other parameters}

Let us consider the case of spherically symmetric fluid distributions
(see Guilfoyle \cite{guilfoyle} for solutions in $\rm d=4$). The
metric is written as
\beq
ds^2 = -W(r)^2 dt^2 + U(r)^2 dr^2 + r^2 d\Omega_{\rm d-2} \, ,
\eeq
where $r$ is the radial coordinate in $(\rm d-1)$ spatial dimensions,
$W$ and $U$ are function of $r$ only, and $ d\Omega_{\rm d -2}$ is the
metric of the unit sphere $\mathbf{S}^{\rm d-2}$. The charged pressure
fluid is bounded by a spherical surface of radius $r=\mathfrak{a}$,
and for $r>\mathfrak{a}$ the metric is given by Tangherlini solution 
\cite{tangherlini}
\beq
ds^2 =
-\left( 1 -\frac{2G_{\rm d}}{\rm d-3} \frac{m}{r^{\rm d-3 }}+
\frac{G_{\rm d}}{({\rm d}-3)^2} \frac{q^2}{r^{2({\rm d} -3)}}\right)
 dt^2 +\frac{dr^2}{1 -\frac{2G_{\rm d}}{\rm d-3} \frac{m}{r^{\rm d-3 }}+
\frac{G_{\rm d}}{({\rm d}-3)^2} \frac{q^2}{r^{2({\rm d} -3)}}} + r^2
d\Omega_{\rm d -2}, \label{tangherliniST}
\eeq
Furthermore, in the asymptotic region the electric potential is given by
\beq
\phi=\frac{1}{\rm d-3}\, \frac{q}{r^{\,\rm d-3}}+\phi_0\, ,
\label{phi-tangherlini}
\eeq
$\phi_0$ being an arbitrary constant which defines the zero of the electric
potential. Notice that no Weyl type relation of any kind is imposed to
the  potentials in the exterior region.

Our aim here is to find the explicit relation between $q$ and $m$. This can
be done, for instance, by direct integration of Eq.~\eqref{mpconditionGH1}.
However, in such a case we need to know the explicit form of the solution.
Alternatively, such a relation can be obtained considering appropriate
junction conditions on the boundary surface $r=\mathfrak a$. In fact, in
the spherically symmetric case, we can integrate Maxwell equation
\eqref{maxGH2}, which furnishes
\beq
Q(r) =  r^{\rm d-2}\,\frac{\phi^\prime}{W \,U}\, ,
\label{qspherical}
\eeq
where the prime denotes the derivative with respect to the radial
coordinate $r$. We then use the Weyl-Guilfoyle relation to write $\phi$ in
terms of $W$,
\beq
\sqrt{G_{\rm d}}\phi = \epsilon\,b-
\epsilon\,\sqrt{\frac{W^2}{a} - \frac{c}{a}} \, ,
\eeq
where $\epsilon=\pm 1$.
With this, Eq.~\eqref{qspherical} reads
\beq
Q(r) = \epsilon\,\frac{1}{\sqrt{G_{\rm d}}}\,\frac{ r^{\rm d-2}\,W^\prime}
{\,U\sqrt{a{W^2} - a{c}}} \, .
\label{qspherical2}
\eeq
Now we use the fact that $Q(r=\mathfrak a)= q$, and that the continuity
of the metric coefficients on the surface $r=\mathfrak a$ implies in
$W(\mathfrak a)^2 = 1/U(\mathfrak a)^2=  1 -\frac{2G_{\rm d}}{\rm d-3}
\frac{m}{\mathfrak a^{\rm d-3 }}+ \frac{G_{\rm d}}{({\rm d}-3)^2}
\frac{q^2}{\mathfrak a^{2({\rm d} -3)}}$. With this, Eq.
\eqref{qspherical2} yields
\beq
ac = a\left[1-\frac{G_{\rm d}\,m^2}{q^2}\right]+
\left(a-1\right) \left(\frac{m}{q}-
\frac{q}{({\rm d}-3)\,\mathfrak a^{\rm d-3}}\right)^2
{G_{\rm d}}\  \, ,  \label{guilfmass}
\eeq
which for $\rm d=4 $ coincides with Eq.~(25) of  \cite{guilfoyle}
and holds for all spherically symmetric charged pressure fluid
distribution whose boundary is the spherical surface of radius
$r=\mathfrak a$.  The proportionality between $q$ and $m$ is recovered
whenever $a=1$, and, moreover, if one imposes further that $c=1- b^2$
it gives $\sqrt{1-c\,}\,q= b\, q= \epsilon\,
\sqrt{G_{\rm d}\,}\, m$.
From Eq.~(\ref{guilfmass}) one sees there is also the possibility $m\,
\mathfrak a^{\rm d-3}=q^2/({\rm d}-3)$, for $a\neq 1$, and still holding
$q\propto m$, more precisely $\sqrt{1-c\,}\,q=\epsilon\sqrt{G_{\rm
d}}\,m$.  The extremal relation $q= \sqrt{G_{\rm d}}\,m$ holds when
the relation among $W^2$ and $\phi$ is a perfect square, for which the
parameter $c$ must vanish, $c=0$, besides $a=1$, implying also in $b= 1.$

\subsubsection{Relation between the mass and charge for a
fluid distribution with no symmetry a priori,
linked to a vacuum that asymptotes the Tangherlini solution}

Let us assume that the charge fluid distribution is such that the
spacetime is asymptotically Tangherlini. This is the case, e. g., of
sources of finite extent with a boundary, or of fluid distributions
without a boundary in which the fluid quantities vanish smoothly in
the asymptotic region in a sufficiently fast way. For these kind of
spacetimes, the metric in the asymptotic region can be taken as the
Tangherlini metric \eqref{tangherliniST}, and the electric potential
is given by Eq.~\eqref{phi-tangherlini}.

Assuming that in the asymptotic region the relation between the
gravitational potential $W$ and the electric potential $\phi$ is of
Weyl-Guilfoyle type (cf. Eq.~\eqref{weylrelationGH1}), $ W^2 = \bar a \,
\left( -\epsilon\,\sqrt{G_{\rm d}}\,\phi+ \bar b\right)^2 + \bar c$,
where a bar here means we are working in the asymptotic region, we
find the conditions
\beq
\bar a = 1, \quad \bar a\left( -\epsilon\,\sqrt{G_{\rm d}}\,\phi_0 +
\bar b\right)q =\epsilon\,\sqrt{G_{\rm d}}\, m, \quad \bar a
\left(-\epsilon\,\sqrt{G_{\rm d}}\,\phi_0 + \bar b\right)^2 + \bar c=
\frac{1}{\bar a}\, \frac{ G_{\rm d}m^2}{q^2} + \bar c =1\,.
\eeq
From these results an interesting conclusion can be drawn:  If it is
assumed that a Weyl-Guilfoyle relation between potentials holds throughout
an asymptotically Tangherlini spacetime then the resulting relation is in
fact a Weyl relation, $ W^2 =\left(-\epsilon\sqrt{G_{\rm
d}}\,\phi+b\right)^2 + c$. A further simplification can be done considering
that the arbitrary constant $\phi_0$ in the electric potential can be put
to zero without loss of generality. With  $\phi_0=0$ we find $\bar b ^2
+\bar c=1$, and therefore
\beq
\bar b\, q=\epsilon\,\sqrt{G_{\rm d}}\,m\, . \label{q2mratio0}
\eeq

That the charge to mass relation is given by Eq.~(\ref{q2mratio0})
in a four-dimensional asymptotically Reissner-Nordstr\"om
spacetime of Weyl type was first found by Gautreau and
Hoffman \cite{gautreau}, using a formal method.
Inspired in this work \cite{gautreau}, let us
see how it can be generalized to $\rm d$-dimensional asymptotically
Tangherlini spacetimes of Weyl-Guilfoyle type.
Following closely Gautreau and Hoffman \cite{gautreau} we assume
there is a fluid distribution of finite extent,
not necessarily spherically symmetric, linked
to a vacuum that faraway asymptotes to the Tangherlini solution,
and moreover, the parameters $a$, $b$, and $c$ defining the
Weyl-Guilfoyle relation inside the fluid are the same
as the parameters of the vacuum region
$\bar a$, $\bar b$, and $\bar c$, i.e.,
$a=\bar a$, $b=\bar b$, and $c=\bar c$. Since in the vacuum region
$\bar a=1$, one also has $a=1$.
Take now the quantity
$\displaystyle{
\nabla_i\left(\frac{1}{W}\,\nabla^i W^2\right) -a
\nabla_i\left(\frac{1}{W}\,\nabla^i\phi^2\right)}$
with $a=1$ (see Eqs.~\eqref{tteinGH2} and \eqref{maxGH2}),
integrate it over a volume ${\cal V}$
and use the Gauss theorem to find
\beqa
\int_{{\cal V}}\left[
\nabla_i\left(\frac{1}{W}\,\nabla^i W^2\right) -
\nabla_i\left(\frac{1}{W}\,\nabla^i\phi^2\right)\right]\, d{\cal
V}= \int_{{\cal S}} \left(\frac{1}{W}\nabla_i\left[W^2 -
G_{\rm d}\phi^2 \right]\right)n^i d{\cal S}\, , \label{massdef01}
\eeqa
where, as before, $d{\cal V}$ is the invariant volume element of the ($\rm
d-1$)-dimensional spatial section ($t=$ const.) of the spacetime, ${\cal
S}$ is the boundary of ${\cal V}$ and $n^i$ is the unit vector orthogonal
to the hypersurface ${ \cal S}$ pointing outwards. Now we take $\cal S$ at
infinity and denote ${\cal S} ={\cal S}_\infty$, so that $\cal V$ stands
for the whole space volume $\cal V_\infty$, and it results that the surface
integration on the right-hand side of Eq.~\eqref{massdef01} is done at
spatial infinity.
Since the spacetime is asymptotically flat, the metric on ${\cal S}_\infty$
can be taken as the Tangherlini metric \eqref{tangherliniST}, with the
electric potential given by Eq.~\eqref{phi-tangherlini}. Under these
conditions the right-hand side of Eq.~\eqref{massdef01} gives
$2\,S_{\rm d-2}G_{\rm d} m$, i.e.,
 \beq
\int_{{\cal V}_\infty}\left[
\nabla_i\left(\frac{1}{W}\,\nabla^i W^2\right) -
\nabla_i\left(\frac{1}{W}\,\nabla^i\phi^2\right)\right]\, d{\cal
V}= \int_{{\cal S}_\infty} \left(\frac{1}{W}\nabla_i\left[W^2 -
G_{\rm d}\phi^2 \right]\right)n^i d{\cal S} =
2G_{\rm d} m\, . \label{massdef01a}
\eeq
Now the field equations Eqs.~\eqref{tteinGH2} and \eqref{maxGH2} are
used to write the first integral on left-hand-side of
Eq.~\eqref{massdef01a} in terms of the fluid quantities and of the
electromagnetic energy density $\rho_{\rm em}$. In the case where the
fluid distribution has no boundary the analysis is straightforwardly
performed and it results 
$ \int_{{\cal V}_\infty}\left[\nabla_i\left(\frac{1}{W}\,\nabla^i
W^2\right) -\nabla_i\left(\frac{1}{W}\,\nabla^i\phi^2\right)\right]\,
d{\cal V} = 2 S_{\rm d-2}G_{\rm d} \int_{{\cal
V}_\infty}\left(\rho_{\rm m}+ \frac{{\rm d}-1}{{\rm d}-3}\,p
+\frac{\phi}{W}\, \rho_{\rm e} \right)W\,d{\cal V}$.
On the other hand, if the source is of finite extent with a boundary,
the spacetime has two distinct regions: The interior region, bounded
by the surface ${\cal S}_{\rm b}$ and whose volume we denote by ${\cal
V}_{\rm b}$; and the exterior region, whose volume we denote by
$\Delta \cal V$, so that we obtain $ \int_{{\cal
V}_\infty}\left[\nabla_i\left(\frac{1}{W}\,\nabla^i W^2\right) -
\nabla_i\left(\frac{1}{W}\,\nabla^i\phi^2\right)\right]\, d{\cal V}= 2
S_{\rm d-2}G_{\rm d} \int_{{\cal V}_{\rm b}}\left(\rho_{\rm m}+
\frac{{\rm d}-1}{{\rm d}-3}\,p +\frac{\phi}{W}\, \rho_{\rm e}
\right)W\,d{\cal V} $.
Hence, using this result and Eq.~\eqref{massdef01a} we finally obtain
\beq
m =  \int_{{\cal V}}\left[\left(\rho_{\rm m}+ \frac{{\rm d}-1}{{\rm
d}-3}\,p\right)W + \phi\,\rho_{\rm e} \right] d{\cal V} \, ,
\label{massdef04}
\eeq
where $\cal V$  stands for the whole volume of the fluid
distribution. The parameter $m$ is the total gravitational mass of the
spacetime.

We now turn attention to the charge definition, for which one can
always make use of the Maxwell equations. Hence, integrating
Eq.~\eqref{maxGH1} over the volume of the source and using the Gauss
theorem it is found
\beq
 \int_{\cal S} \frac{1}{W}\nabla_i\phi\, n^i d{\cal S} =- S_{\rm d-2}\,
\int_{\cal V_S} \rho_{\rm e}\,d{\cal V} \, , \label{maxGHint}
\eeq
where $\cal S$ is the hypersurface bounding the volume $ {\cal V_S}$.
Again the left-hand side integration can be done over an infinitely large
spherical surface, with the metric on $\cal S$
being the Tangherlini metric \eqref{tangherliniST}, and with the electric
potential being given by Eq.~(\ref{phi-tangherlini}).
Then, the left-hand side of Eq.~\eqref{maxGHint} gives  $-\,S_{\rm d-2} q$.
We may then identify the quantity on the right-hand-side of
Eq.~\eqref{maxGHint} to the total charge of the source $q$,
\beq
q = \int_{\cal V_S} \rho_{\rm e}\,d{\cal V}=\int_{\cal V_\infty} \rho_{\rm
e}\,d{\cal V}\, .\label{totalcharge}
\eeq
Finally, the constraint \eqref{mpconditionGH1} implies that the total
charge of a spacetime containing a charged pressure fluid
distribution of finite size satisfying the conditions of Theorem
\ref{theoGH1-rel} is proportional to the ADM mass of the
spacetime. This can be shown as follows: Integrating
Eq.~\eqref{mpconditionGH1} over the whole volume of the source, ${\cal
V}_{\rm b}$, and using Eqs.~\eqref{massdef04} and \eqref{totalcharge}, we
find
\beq
 b\, q = \epsilon\,\sqrt{G_{\rm d}}\, m  \, , \label{q2mratio}
\eeq
which is the same as Eq.~\eqref{q2mratio0}, since we are assuming $b =\bar
b$.  This result shows that for a compact fluid distribution with no
symmetry a priori, in which the spacetime tends asymptotically to the
Tangherlini solution, with the potentials $W$ and $\phi$ being related by
$W^2 = a\left(-\epsilon\,\sqrt{G_{\rm d}}\,\phi+ b\right)^2 + c$, and with
the mass density $\rho_{\rm m}$, the pressure $p$, the charge density
$\rho_{\rm e}$ being related by $ b\rho_{\rm e}=\epsilon\,\sqrt{G_{\rm
d}}\, \left[\left( \rho_{\rm m}+ \frac{\rm d-3} {\rm d-1} \,p\right)W
+\phi\rho_{\rm e}\right]$, the total mass and the total electric
charge of the spacetime are proportional to each other with the
proportionality constant being exactly $b$. For a fluid mass distribution
with spherical symmetry, the relation \eqref{q2mratio0} can be derived
directly from Eq.~\eqref{guilfmass}. Indeed, under the Gautreau-Hoffman
assumptions, one has $b=\bar b$, and thus $\bar b^2{q^2}=G_{\rm d} m^2$, so
that one finds immediately from Eq.~\eqref{guilfmass}, that $a=1$ and $b^2
+c =1$ is a solution. So the Gautreau-Hoffman mass relation is, in the
spherically symmetric case, a particular instance of the Guilfoyle
relation.

We can also show that the integral given by Eq.~\eqref{massdef04} is
the total gravitational energy, i.e., the total mass, of a source and
is equal to the Tolman mass \cite{tolman}. Using Tolman formula we
find
\beq
M =
\int_{{\cal V_\infty}} \left(\rho_{\rm m}+ \frac{{\rm d}-1}{{\rm
d}-3}\,p+\rho_{\rm em}\right)W\, d{\cal V},
\label{tolmanmass}
\eeq
where we assumed that the charged fluid distribution is of finite extent.
Using Eqs.~\eqref{tolmanmass} and \eqref{massdef04} we get
\beq \label{massrelation0}
M - m = \int_{{\cal V}_{\rm b}} \left(W \rho_{\rm em} -\phi\,\rho_{\rm e}
\right)d{\cal V} +\int_{\Delta{\cal V}} W\,\rho_{\rm em}\, d{\cal V}\, ,
\eeq
where $\Delta \cal V$ stands for the volume outside the matter
distribution. We can get rid of the term containing the product $\phi\
\rho_{\rm e}$ on the right-hand side of Eq.~\eqref{q2mratio}.  
Multiplying Eq.~\eqref{maxGH1} by $\phi$, using the identity, $\phi
\nabla_i \left(\frac{1}{W}\,\nabla^i\phi\right)= \nabla_i
\left(\frac{\phi}{W}\,\nabla^i\phi\right) -
\left(\nabla^i\phi\right)^2/W$, integrating over a space volume ${\cal
V_S}$ and using the Gauss theorem, it is obtained that  $\displaystyle{
\int_{\cal S} \left(\frac{\phi}{W}\,\nabla_i\phi\right)\, n^i d{\cal
S}- \int_{{\cal V_S}}\frac{1}{W}\left(\nabla^i\phi\right)^2\,\,d{\cal
V} = - S_{\rm d-2}\int_{{\cal V_S}}\phi\, \rho_{\rm e} \, d{\cal V}}$.
Then, using Eq.~\eqref{emdensity} to bring in the electromagnetic
energy density, $ \left(\nabla^i\phi\right)^2/W= S_{\rm d-2} W\,
\rho_{\rm em}$, it is finally found $S_{\rm d-2}\int_{{\cal V}_{\rm
b}}\left(\phi\,\rho_{\rm e} -W\rho_{\rm em}\right)\, d{\cal V} =
-\int_{{\cal S}_{\rm b}} \left(\frac{\phi}{W}\,\nabla_i\phi\right)\,
n^i d{\cal S} \, ,$ where ${\cal S}_{\rm b}$ is the boundary surface
of the fluid distribution.  Substituting this result into
Eq.~\eqref{massrelation0} it gives
\beq
M - m = \frac{1}{S_{\rm d-2}}\int_{{\cal S}_{\rm b}}
\left(\frac{\phi}{W}\,\nabla_i\phi\right)\, n^i d{\cal S}
 +\int_{\Delta{\cal V}} \rho_{\rm em} W\, d{\cal V}, \label{massesrelation}
\eeq
where ${\cal S}_{\rm b}$ is the surface which bounds the source. It
seems that the mass $m$ is in general different from the Tolman mass
$M$.  However, as verified above, for $\rm d$-dimensional spherically
symmetric systems of Weyl-Guilfoyle type it results $m=M$. In fact,
one can show that under certain conditions the two integral terms on
the right-hand side of Eq.~\eqref{massesrelation} cancel each other
out. For we use the vacuum field equations and the Gauss theorem to
transform the volume integration over $\Delta{\cal V}$ into a surface
integration, viz, $\displaystyle{ S_{\rm d-2} \int_{\Delta {\cal V}}
\rho_{\rm em} W\, d{\cal V} = \int_{{\cal S}_\Delta}
\left(\frac{\phi}{W}\,\nabla_i\phi\right)\, \bar n^i d{\cal S}} $,
where ${\cal S}_\Delta$ is the (closed) surface boundary to the volume
$\Delta {\cal V}$, and $\bar n^i$ stands for the unit vector
orthogonal to ${\cal S}_\Delta$ pointing outwards. Note that the
boundary $S_\Delta$ is composed of two closed surfaces. The external
boundary ${\cal S}_\infty$, and inner boundary, which coincides with
the boundary of the source, ${\cal S}_{\rm b}$. The unit vector $\bar
n^i$ on ${\cal S}_{\rm b}$ pointing outwards with respect to
$\Delta{\cal V}$ is parallel to $ n^i$ but points inwards with respect
to ${\cal V}_{\rm b}$, namely, $\bar n^i = - n^i$ where $n^i$ is the
same as in Eq.~\eqref{massesrelation}.
Therefore, we have the identity $\displaystyle{\int_{{\cal S}_\Delta}
\left(\frac{\phi}{W}\,\nabla_i\phi\right)\, \bar n^i d{\cal S} = -
\int_{{\cal S}_{\rm b}}\left(\frac{\phi}{W}\,\nabla_i\phi\right)\, n^i 
d{\cal  S}+  \int_{{\cal S}_\infty}
\left(\frac{\phi}{W}\,\nabla_i\phi\right)\, \bar n^i d{\cal S}} $.
Since the source is of finite extent the spacetime is asymptotically
Tangherlini, cf. Eqs.~\eqref{tangherliniST} and
\eqref{phi-tangherlini}, the integral over ${\cal S}_\infty$ vanishes and,
after substituting the result into Eq.~\eqref{massesrelation}, we find
\beq
M=m\,, \label{equalmasses}
\eeq
i.e., the total mass is the Tolman mass \cite{tolman} 
(for the Tolman mass and mass in charged matter see
\cite{whittaker,florides1,florides2}, and also \cite{cohen2}).

\section{Further comments and conclusions}
\label{conclusions}

We have studied the structure of the sources produced by Weyl type
systems, including systems obeying a Weyl-Guilfoyle relation, both in
the Newton-Coulomb theory with matter in $\rm d-1$ space dimensions
and in the Einstein-Maxwell theory with matter in $\rm d$ spacetime
dimensions.

In the Newton-Coulomb case, we have rendered theorems by Bonnor for
charged dust fluids into higher dimensions, and obtained new results
for charged pressure fluids. For zero pressure fluids, it follows that
the gravitational potential $V$ is a function of the electric
potential $\phi$ alone, $V = V(\phi)$. In the case of a nonzero
pressure fluid, the equations of fluid dynamics in Eulerian
description together with the Poisson equations for the potentials
with the assumption of a Weyl type ansatz for the potentials, $V =
V(\phi)$, implies that the pressure is also a function of $\phi$
alone, and then all fluid quantities are given in terms of the
gravitational potential. If one assumes further that the relation
between the pressure gradient $dp/d\phi$ is proportional to the charge
density, then the gravitational potential is given by $V(\phi)=
-\epsilon\,\beta\sqrt{G_{\rm d}}\,\phi +\gamma$, with $\beta$ and
$\gamma$ being arbitrary constants and $\epsilon=\pm 1$, with the
matter and charge densities being proportional to each other too.

In the case of the relativistic theory things are more intricate and
more interesting. First, with the Weyl ansatz $W = W(\phi)$, the
d-dimensional Einstein-Maxwell theory in vacuum yields the Weyl
quadratic relation between the metric potential $W^2$ and the electric
gauge potential $\phi$.  Then, a series of results for
the Einstein-Maxwell with matter theory for fields of Weyl type in four
spacetime dimensions can be rendered into higher dimensional
spacetimes. The most important new result of our analysis is the
generalization of the Gautreau-Hoffman relation among the fluid
quantities when the potentials are related through the Weyl-Guilfoyle
relation \eqref{weylrelationGH1}. This analysis, done in
Sect.~\ref{newthm-rel}, shows that the most general charged pressure
fluid in which the metric gravitational potential $W$ satisfying the
Weyl-Guilfoyle quadratic relation $W^2 =
a\,\left(-\epsilon\sqrt{G_{\rm d}}\,\phi+b\right)^2+c, $ with
constants $a$, $b$, and $c$, corresponds to different systems when
compared to the original Weyl quadratic relation as
Eq.~\eqref{weyloriginalrelation}, in which $a=1$. To see that, consider
the following reparameterization of the fields $W \longrightarrow
\alpha_0\, W \, , \phi \longrightarrow \, \alpha_1\phi\, ,$ with
constant $\alpha_0$ and $\alpha_1$. Taking these transformations into
the system of equations given by Eqs.~\eqref{ttein}, \eqref{max2} and
\eqref{conserveq}, we can conclude that the Einstein-Maxwell system of
equations is invariant only if $\alpha_0=\alpha_1$, as expected (this
corresponds to a rescaling of the time coordinate, $t\longrightarrow
t/{\alpha_0}$). Therefore, any rescaling of the potentials $W$ and
$\phi $ for which $\alpha_0\neq \alpha_1$ leads to a different
system. Hence, a new relation among fluid quantities and the
electromagnetic energy density is found in four and higher dimensions,
cf. Eq.~\eqref{mpconditionGH1}.  Upon connection of an interior
charged solution to an exterior Tangherlini solution, we found a
relation between the mass, the charge and the several quantities of
the interior solution. It was also shown that for sources of finite
extent the mass is identical to the Tolman mass.


\section*{Acknowledgments}
We thank Antares Kleber for many conversations. 
We thank Observat\'orio Nacional of Rio de Janeiro for hospitality
while part of the present work was being done. This work was partially
funded by Funda\c c\~ao para a Ci\^encia e Tecnologia (FCT) -
Portugal, through project PPCDT/FIS/57552/2004.  VTZ thanks Funda\c
c\~ao de Amparo \`a Pesquisa do Estado de S\~ao Paulo (FAPESP)
(through project 2007/04278-2) and Conselho Nacional de
Desenvolvimento Cient\'\i fico e Tecnol\'ogico of Brazil (CNPq) for
financial help.

\end{document}